\documentclass[letterpaper]{article} 
\usepackage{aaai24arxiv}
\usepackage{times}  
\usepackage{helvet}  
\usepackage{courier}  
\usepackage[hyphens]{url}  
\usepackage{graphicx} 
\usepackage{natbib}  
\usepackage{caption} 
\usepackage{amsmath}
\usepackage{amssymb}
\usepackage{amsthm}
\usepackage{stmaryrd}
\usepackage{xcolor}
\usepackage{booktabs}
\usepackage{algorithm2e}
\usepackage{float}
\usepackage{nicefrac}
\usepackage{dsfont}

\newcommand{\Benjamin}[1]{\noindent\textcolor{blue}{$\langle$B: \textsf{#1}$\rangle$}}

\newtheorem{lemma}{Lemma}

\newcommand{\EE}{\mathbb{E}}

\newcommand{\RR}{\mathbb{R}}
\newcommand{\T}{{\text{\tiny\sffamily\upshape\mdseries T}}}
\def\dd{\mathrm{d}}
\def\by{\textbf{y}}
\def\bY{\textbf{Y}}
\def\bmu{\boldsymbol{\mu}}
\def\bSigma{\boldsymbol{\Sigma}}
\def\Ygk{\mathbf{Y}_{g,k}}
\def\Var{\text{Var}}

\usepackage{tikz}
\usepackage{xcolor}
\definecolor{iblue}{RGB}{0,0,156}
\definecolor{ired}{RGB}{192,0,0}
\definecolor{igrey}{RGB}{108,123,139}

\usepackage[pagebackref=true]{hyperref}
\hypersetup{
	colorlinks=true,
	urlcolor=iblue,
	citecolor=iblue,
	linktoc=all,
	linkcolor=iblue
	}

\usepackage{newfloat}
\usepackage{listings}
\DeclareCaptionStyle{ruled}{labelfont=normalfont,labelsep=colon,strut=off} 
\floatstyle{ruled}
\newfloat{listing}{tb}{lst}{}
\floatname{listing}{Listing}
\title{Maximizing the Success Probability of Policy Allocations
in Online Systems}
\author{
    Artem Betlei, \textsuperscript{\rm 1} 
    Mariia Vladimirova, \textsuperscript{\rm 1} 
    Mehdi Sebbar, \textsuperscript{\rm 2} \\
    Nicolas Urien, \textsuperscript{\rm 2} 
    Thibaud Rahier, \textsuperscript{\rm 1} 
    Benjamin Heymann \textsuperscript{\rm 1}
}
\affiliations{
    \textsuperscript{\rm 1} Criteo AI Lab, France\\
    \textsuperscript{\rm 2} Criteo Ad Landscape, France

    \{a.betlei, m.vladimirova, m.sebbar, n.urien, t.rahier, b.heymann\}@criteo.com
}

\usepackage{bibentry}

\begin{document}

\maketitle

\begin{abstract}
    The effectiveness of advertising in e-commerce largely depends on the ability of merchants to bid on and win impressions for their targeted users. 
    The bidding procedure is highly complex due to various factors such as market competition, user behavior, and the diverse objectives of advertisers.
    In this paper we
    consider the problem at the level of user timelines instead of individual bid requests, manipulating full policies (i.e. pre-defined bidding strategies) and not bid values. 
    In order to optimally allocate policies to users, typical multiple treatments allocation methods solve knapsack-like problems which aim at maximizing an expected value under constraints.
    In the industrial contexts such as online advertising, we argue that optimizing for the probability of success is a more suited objective than expected value maximization, and we introduce the \texttt{SuccessProbaMax} algorithm that aims at finding the policy allocation which is the most likely to outperform a fixed reference policy.
    Finally, we conduct comprehensive experiments
    both on synthetic and real-world data 
    to evaluate its performance. The results demonstrate that our proposed algorithm outperforms conventional expected-value maximization algorithms in terms of success rate.
\end{abstract}

\section{Introduction}
Optimizing marketing effectiveness relies on using individualized bidding policies, exploiting the fact that each user responds differently. A \textit{policy} may include a set of rules or actions over an extended period of time, e.g., cash bonuses, promotion  and display ad shown to  consumers on online platforms. Without loss of generality, we take the narrow view of bidding for display advertising in order to ground our research into a real life application.  In this context,  the task at hand is to specify a full bidding strategy (the policy) on the future advertisement opportunities for each given users during a given time period.

In practice, it is typical to have a fixed budget allocated to a campaign.
From an advertising perspective, a bidding strategy must maximize the total expected revenue while ensuring that the expected total cost does not exceed a specified budget.

Usually, this problem is modeled as a multiple choice knapsack problem~\citep{demirovic2019investigation,zhou2023direct} with the objective to select at most one item (bid policy) from each user such that the sum of the weights (expected cost) of selected items does not exceed the capacity (budget) while the total \textit{reward} (expected revenue) is maximized. This problem is known to be NP-hard, although it can be tackled with mixed integer linear programming or through Lagrangian relaxation~\citep{sinha1979multiple}.

From a causal perspective, it is classical to consider every individual ad as a treatment, and the optimization problem goal is to maximize the total causal effect of these treatments by correctly assigning treatments to users. There exist various approaches for individual treatment assignment that differ by the objective function they optimize: learning models to predict either outcomes, causal effects or directly the optimal treatment assignment. \citet{fernandez2022comparison} compare these approaches analytically and show that the assignment learners optimize the bias-variance tradeoff with respect to decision-making errors. 

Optimization at the opportunity \textendash or bid \textendash level, which we refer as \textit{bid by bid} optimization, requires to attribute each observed reward to the action that actually caused it, e.g. each conversion must be attributed to a shown ad. This attribution problem is very complex as there usually are several ads displayed in the few hours preceding each conversion~\citep{10.1145/3447548.3467280,bompaire2021robust,Ji_Wang_2017,10.1145/2351356.2351363}. It causes fundamental problems in the estimation of the causal effects and makes the bid by bid optimization extremely difficult in practice. 

Furthermore, display advertising campaigns, like many other online systems, are operated under several business and technical constraints. 
In particular, it is typical for an advertising campaign to have a budget constraint.
Several algorithms allow adapting  bid by bid optimization techniques to  such constraints~\citep{castiglioni2022unifying,doi:10.1287/opre.2021.2167}. 
While these algorithms have their merits and are largely deployed in practice, they are, 
however,  poorly suited for \textit{causal} bid by bid methods. This is because (a) typical causal methods inherently  suppose the absence of causal interaction between the treatment units --- such assumption is in general  violated when mixing causal method for bid by bid optimization and budget pacing; (b) the overall methodology needs to trade off marginal value and marginal  future total cost~\citep{10.1145/3447548.3467280}, which is arguably intractable at the bid level.

Our first idea is to reformulate the problem at the user timeline level (i.e. considering all the bid requests and subsequent events relative to a user along a given time period) which implies to consider entire policies instead of individual bids.
With this new formulation, the optimal policy allocation search is framed as a multiple treatment allocation problem, and the causal effects (cost and value) of policies are much easier to estimate than that of individual bids. Our approach is not to be understood as in competition with usual bid by bid design approaches \citep{moriwaki2021real} but rather complementary. Indeed, any bid by bid design approach could be included as one of the candidate policies we wish to choose from when allocating policies to users with our methodology. If a bid by bid design policy happens to be globally optimal, our method will simply conclude that the optimal policy allocation consists in assigning this policy to every user.

However, we claim that searching for the policy allocation function which maximizes an expected value under an expected cost constraint (which is typically done in treatment allocation problems) is not always the best objective.
In a large organization, it is often necessary to have  guidelines that allow for consistent decision-making regarding product design and improvements.
Without such guidelines, individuals cannot handle trade-offs between different quantities (for example, quality and volume) consistently across the whole organization. One may think about designing medication (which should be efficient but also avoid negative side-effects) or electrical batteries (which should have a big enough capacity while not relying too much on rare materials). This leads to the definition of a \textit{success} across organizations, e.g. in online advertising, it corresponds to increasing generated value \textit{without increasing the cost} with respect to a reference outcome. 
Taking this as a premise, the (constrained) maximization of a single quantity \textendash such as revenue \textendash is not anymore the right criterion as it does not account for the uncertainty underlying the phenomenon at play,
nor does it account for what will be considered a success. 

This motivates the focus on finding the policy allocation resulting in the \textit{highest probability of success}.
While every metric has its pros and cons, we believe a focus on success probability, with a very flexible notion of \textit{success}, is of particular operational interest, see Fig.~\ref{fig:sketch_spm} for an illustrative example (we refer to Section~\ref{sec:success-proba-max} for a detailed description). 

In summary, this work presents the following contributions:
\begin{itemize}
    \item 
    We formally propose the idea of framing the optimization problem at the policy level instead of focusing on bid by bid design, and mathematically formalize both the expected value maximization and the success probability maximization problems.
    \item 
    We develop a novel customized solution to address the specificities of the success probability optimization problem.
    \item Finally, we present a series of numerical experiments which were conducted on both synthetic and real-world data, showing that our approach outperforms traditional value maximization methods in terms of success rate guarantees.
\end{itemize}

\tikzset{every picture/.style={line width=0.75pt}} 
\begin{figure}[h]
  \begin{center}
\begin{tikzpicture}[x=1.25pt,y=1.25pt,yscale=-1,xscale=1]

\draw  [draw opacity=0][fill={rgb, 255:red, 208; green, 2; blue, 27 }  ,fill opacity=0.2 ] (245.7,33.61) -- (286,33.61) -- (286,117.91) -- (245.7,117.91) -- cycle ;
\draw  [draw opacity=0][fill={rgb, 255:red, 184; green, 233; blue, 134 }  ,fill opacity=0.3 ] (160.7,32.91) -- (245.7,32.91) -- (245.7,117.91) -- (160.7,117.91) -- cycle ;
\draw  (160,117.91) -- (286,117.91)(246,34.72) -- (246,133.22) (279,112.91) -- (286,117.91) -- (279,122.91) (241,41.72) -- (246,34.72) -- (251,41.72)  ;
\draw  [color={rgb, 255:red, 245; green, 166; blue, 35 }  ,draw opacity=1 ][fill={rgb, 255:red, 245; green, 166; blue, 35 }  ,fill opacity=0.3 ] (200,84.1) .. controls (204.19,71.32) and (212.39,62.54) .. (218.3,64.49) .. controls (224.21,66.43) and (225.6,78.35) .. (221.41,91.13) .. controls (217.21,103.9) and (209.02,112.68) .. (203.11,110.74) .. controls (197.2,108.8) and (195.8,96.87) .. (200,84.1) -- cycle ;
\draw  [draw opacity=0][fill={rgb, 255:red, 208; green, 2; blue, 27 }  ,fill opacity=0.2 ] (160,117.91) -- (286,117.91) -- (286,132.5) -- (160,132.5) -- cycle ;
\draw  [draw opacity=0][line width=3] [line join = round][line cap = round] (292,69) .. controls (292,69) and (292,69) .. (292,69) ;
\draw  [color={rgb, 255:red, 245; green, 166; blue, 35 }  ,draw opacity=1 ][line width=3] [line join = round][line cap = round] (210.8,87.57) .. controls (210.8,87.57) and (210.8,87.57) .. (210.8,87.58) ;
\draw  [color={rgb, 255:red, 0; green, 0; blue, 0 }  ,draw opacity=1 ][line width=3] [line join = round][line cap = round] (246.18,117.94) .. controls (246.18,117.94) and (246.18,117.95) .. (246.18,117.95) ;
\draw  [color={rgb, 255:red, 74; green, 144; blue, 226 }  ,draw opacity=1 ][fill={rgb, 255:red, 74; green, 144; blue, 226 }  ,fill opacity=0.3 ] (234.85,72.1) .. controls (239.04,59.32) and (247.23,50.54) .. (253.14,52.49) .. controls (259.05,54.43) and (260.45,66.35) .. (256.25,79.13) .. controls (252.06,91.9) and (243.86,100.68) .. (237.95,98.74) .. controls (232.04,96.8) and (230.65,84.87) .. (234.85,72.1) -- cycle ;
\draw  [color={rgb, 255:red, 74; green, 144; blue, 226 }  ,draw opacity=1 ][line width=3] [line join = round][line cap = round] (245.88,74.95) .. controls (245.88,74.95) and (245.88,74.96) .. (245.88,74.96) ;

\draw (256.84,74.22) node [anchor=north west][inner sep=0.75pt]  [color={rgb, 255:red, 74; green, 144; blue, 226 }  ,opacity=1 ,rotate=-0.1] [align=left] {{\tiny sol. to }\\{\tiny eq. (2)}};
\draw (173.66,80.58) node [anchor=north west][inner sep=0.75pt]  [color={rgb, 255:red, 245; green, 166; blue, 35 }  ,opacity=1 ,rotate=-0.1] [align=left] {\tiny sol. to \\ \tiny eq. (3)};
\draw (177,44.4) node [anchor=north west][inner sep=0.75pt]  [color={rgb, 255:red, 65; green, 117; blue, 5 }  ,opacity=1 ]  {$\mathcal{S}_{y_0}$};
\draw (216.07,121.09) node [anchor=north west][inner sep=0.75pt]  [font=\tiny]  {$\left( y_{0}^{c} ,\ y_{0}^{v}\right)$};
\draw (253,35.31) node [anchor=north west][inner sep=0.75pt]  [font=\tiny]  {$y^{v}$};
\draw (275,122.31) node [anchor=north west][inner sep=0.75pt]  [font=\tiny]  {$y^{c}$};
\end{tikzpicture}
  \end{center}
  \vspace{-4mm}
  \caption{Distributions the (cost,value) outcome vector $\mathbf{Y}=(Y^c, Y^v)$. In blue for the solution of (2) (maximization of $\mathbb{E}[Y^v]$ under the condition $\mathbb{E}[Y^c] \leq y^c_0$) ; and in orange the solution of (3) (maximization of the success probability).}
  \label{fig:sketch_spm}
\end{figure}
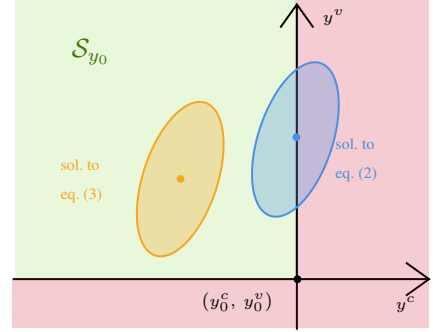

\vspace{-3pt}

\section{Problem formulation}\label{sec:problem-formulation}
\paragraph{Preliminary considerations} Throughout this section, we will implicitly refer to a given time period $\tau$ of length $\Delta t$, i.e. $\tau = [t_0, t_0 + \Delta t]$.
We consider a given advertiser $Adv$ who has a fixed budget $C$ to spend over period $\tau$. 

\paragraph{Set of candidate policies} We assume given a set of $K$ \textit{candidate policies} $\Pi = \{\pi_0, \pi_1, \dots, \pi_{K-1} \}$ each encapsulating \textit{bidding strategies} that may be applied by $Adv$ to each user \textbf{consistently throughout the period $\tau$}. The \textit{reference policy} $\pi_0$ is the default bidding strategy used by $Adv$ (typically corresponding to the strategy which is already rolled out in production for this advertiser). This set of policies $\Pi$ can be thought of as a collection of potential treatments in a \textit{multiple treatment allocation} problem. 
Note that we do not consider treatments at the level of bidding opportunities here, but at the level of an extended time period, during which we apply policies \textendash or bidding strategies \textendash which each have an integrated way to decide on how to bid on each user for all the opportunities that will arise during period $\tau$.

\paragraph{Random variables and potential outcomes}
Considering the above setup, and with respect to any given user $u$ targetable by $Adv$, we define the following random variables: 
\begin{itemize}
    \item $\mathbf{X} \in \mathcal{X} \subset \mathbb{R}^d$ contains a snapshot of the \textit{features} of $u$ captured at time $t_0$,
    \item $\mathbf{Y} = (Y^v, Y^c) \in \mathcal{Y} \subset \mathbb{R}_{+}^2$ contains respectively the \textit{value} generated by $u$ in favor of $Adv$ during period $\tau$ and the \textit{cost} $Adv$ spent to advertize to $u$.
\end{itemize}

For any $\pi\in\Pi$, we denote $\mathbf{Y}(\pi) = \left(Y^v(\pi), Y^c(\pi)\right)$ the \textit{potential outcomes} \cite{rubin1974estimating} we would have observed \textbf{had $\pi$ been applied to $u$} during $\tau$.
In what follows, we consider the tuple $(\mathbf{X}, \mathbf{Y}(\pi_0), \dots, \mathbf{Y}(\pi_{K-1}))$ has an underlying probability distribution $\mathbb{P}$, which we will overload by simplicity to designate its marginals and conditionals. All expectancy notations $\mathbb{E}$ will refer implicitly to $\mathbb{P}$.

\paragraph{Factuals and counterfactuals}
Assuming that we apply policy $\pi_u\in\Pi$ to a given user $u$ during $\tau$, we denote $\mathbf{y}_u = (y^v_u, y^c_u)$ the corresponding realization of the outcome variable $\mathbf{Y}$, and $\{\mathbf{y}_u(\pi)\}_{\pi\in\Pi}$ the corresponding realizations of the potential outcomes variables $\{\mathbf{Y}(\pi)\}_{\pi\in\Pi}$.
In that case, $\mathbf{y}_u = \mathbf{y}_u(\pi_u)$ is called the observed \textit{factual outcome} and the $\{\mathbf{y}_u(\pi_u)\}_{\pi\in\Pi \setminus \{\pi_u\}}$ are the un-observed \textit{counterfactual outcomes}.

\paragraph{Population random variables}
Let $\mathcal{U} = \{1, \dots, N\}$ be the set of users who are targetable by $Adv$ during period $\tau$. We have access to a randomized controlled trial (RCT) \textendash in this case also called an online controlled experiment or A/B test \textendash on population $\mathcal{U}$ during this period, \textbf{randomly} assigning to each of those $N$ users the $K$ potential policies in $\Pi$. 

Formally, let $\{K_u\}_{u\in\mathcal{U}}$ be N i.i.d. uniform categorical variables with values in $\{0, \dots, K-1\}$. Each $u\in\mathcal{U}$ is assigned to the policy $\pi_{K_u}$ during $\tau$, resulting in the definition of the collection $\{(\mathbf{X}_u, \mathbf{Y}_u)\}_{u\in\mathcal{U}}$, where the $(\mathbf{X}_u, \mathbf{Y}_u) = \left(\mathbf{X}_u, \mathbf{Y}_u(\pi_{K_u})\right)$.
We will denote $\mathbb{P}$\footnote{We drop the reference to $\mathcal{U}$ in $\mathbb{P}$ for simplicity.} the probability distribution of $\{(\mathbf{X}_u, \mathbf{Y}_u)\}_{u\in\mathcal{U}}$.
Lastly we assume that, in expectation, $Adv$ exactly spends their advertising cost budget $C$ during $\tau$ had they assigned the default policy $\pi_0$ to every user, i.e. 
$
\mathbb{E}\left[\sum_{u\in\mathcal{U}}Y^c_u(\pi_0)\right]= C.
$

\subsection{Expected value maximization problem}\label{ssec:value-max-problem}
\subsubsection{At the user level}
In the setup we introduced, one aim may be to find an \textit{optimal policy allocation}, i.e. a mapping from $\mathcal{X}$ to policies from $\Pi$ so that \textit{expected total value} generated in favor of $Adv$ is maximized, while respecting (in expectation) the total budget constraint.

Formally, we are looking for a solution $\phi^*:\mathcal{X} \rightarrow \Pi$ to the problem:
\begin{align}
\label{eq:knapsack-general-expectation}
\underset{\phi\in \Pi^\mathcal{X}}{\max} \ \ \mathbb{E}\left[\sum_{u\in\mathcal{U}}Y_u^v(\phi(\mathbf{X}_u))\right]
\ \text{s.t.} \ \mathbb{E}\left[\sum_{u\in\mathcal{U}}Y_u^c(\phi(\mathbf{X}_u))\right] \leq C.
\end{align}

\vspace{-5pt}

In practice, $\Pi^\mathcal{X}$ is very large and hard to explore efficiently, making \eqref{eq:knapsack-general-expectation} a difficult problem, especially since it involves estimation of $K$ potential outcomes in parallel.
A crucial observation is that the problem may be simplified by reducing it to a partition of the space $\mathcal{X}$, which leads to a reparametrization of \eqref{eq:knapsack-general-expectation}, as explained in the next subsection.

\subsubsection{Assuming a given partitioning of the user space}\label{sec:partitioningsetup}

We consider given a partition function $\gamma:\mathcal{X} \rightarrow \mathcal{G}$ where $\mathcal{G} = \{1, \dots, M\}$ contains the indexes of the partition components (or buckets). Reasoning at a bucket-level instead of the user-level is practical in causal estimation setups since it enables to circumvent the fundamental problem of causal inference \citep{betlei2021differentially} and is more compliant with privacy restrictions \citep{kleber2019turtledove}. A reasonable partitioning can be chosen by the domain knowledge or recursively with causal trees through heterogeneous treatment effect estimation
\citep{athey2016recursive,wager2018estimation,tu2021personalized,ai2022lbcf}. 

Given the partition function $\gamma$, we propose to simplify problem \eqref{eq:knapsack-general-expectation}: instead of searching through all $\phi$s in $\Pi^\mathcal{X}$, we restrict our search to the allocations of the form $\psi \circ \gamma$ where $\psi \in \Pi^\mathcal{G}$. In short, we look for allocation functions that assign all users belonging to the same bucket $g\in\mathcal{G}$ to the same policy $\pi\in\Pi$.

Formally, this leads to the reparametrized problem, where we are looking for a solution $\psi^* : \mathcal{G} \rightarrow \Pi$ to the problem:
\begin{align}
\label{eq:knapsack-grouped-expectation}
\underset{\psi\in \Pi^\mathcal{G}}{\max} \ \ \mathbb{E}\left[\sum_{u\in\mathcal{U}}Y_u^v(\psi(G_u))\right] \ 
\text{s.t.} \ \mathbb{E}\left[\sum_{u\in\mathcal{U}}Y_u^c(\psi(G_u))\right] \leq C.
\end{align}

\vspace{-5pt}

where $G_u=\gamma(\mathbf{X}_u)$ for all $u\in\mathcal{U}$.

\subsubsection{Solving the expected value maximization problem}\label{ssec:value-max-solving}
The value expectation maximization problem formalized in \eqref{eq:knapsack-grouped-expectation} may be solved using mixed integer linear programming or Lagrangian relaxation approaches, which make the problem tractable in practice despite being NP-Hard~\citep{sinha1979multiple}. Nevertheless, the knapsack formulation remains a proxy to the marketing problem and its solution does not always align with the business goal. 

\paragraph{Remark 1 \textendash mix of A and B rollout}
If number of buckets $M = 1$, we assign \textbf{all users} to the same policy. This corresponds to a typical rollout decision in an online advertising company: we are A/B testing multiples policies, then depending on the results choosing which one should be rolled out.
Our setup allows for a rollout of a \textbf{mix} of tested policies, given by function $\psi$. 

\paragraph{Remark 2 \textendash relaxing the allocation space}
We can relax problems \eqref{eq:knapsack-general-expectation} \eqref{eq:knapsack-grouped-expectation} by allowing for \textit{soft allocations}, i.e. mappings from $\mathcal{X}$ (resp. $\mathcal{G}$) to $\Delta = \Delta (\Pi)$ where $\Delta$ denotes all categorical distributions with values in $\Pi$:
$$
\Delta := \left\{\left(p(k)\right)_{k\in\llbracket 1, K-1\rrbracket} \in [0,1]^K \ \text{s.t.}\ \sum_k p(k) = 1\right\}.
$$
For $\psi\in \Delta^\mathcal{G}$ and $g\in\mathcal{G}$ and for convenience of notations, we will refer to the $k$th component of $\psi(g)$ \textendash i.e. the probability for $\psi$ to assign a user in bucket $g$ to policy $\pi_k$ \textendash as $\psi(g, k)$.

\subsection{Success probability maximization problem}\label{ssec:proba-max-problem}
In this section, we will focus on the case where we are given a partitioning of the user space and consider more general soft allocation setup presented in \textit{Remark 2} at the end of the previous section.

Instead of searching for the allocation that maximizes the expected value under constraint as in \eqref{eq:knapsack-general-expectation} and \eqref{eq:knapsack-grouped-expectation}, one can also be interested in maximizing their \textit{success probability}, especially in cases where the variance of the variables at play is high. For instance, a policy $\psi^*$ that satisfies \eqref{eq:knapsack-grouped-expectation} might deliver very bad values occasionally. As explained in the introduction, the risk-aversion of industrial players often motivates them to prefer reliable small-increments to uncertain substantial ones.

Instead, we suppose there is an agreement beforehand on the definition of the \textit{success} of a given policy allocation function $\psi: \mathcal{G} \rightarrow \Delta$ through the characterization of a convex region $\mathcal{S}\subset\mathcal{Y}$ such that ``$\psi$ is successful on a the set of users $\mathcal{U}$'' is equivalent to $\sum_{u\in\mathcal{U}}\mathbf{Y}_u(\psi) \in \mathcal{S}$, where for any $\psi \in \Delta^\mathcal{G}$ we denote by simplicity $\mathbf{Y}(\psi) := \mathbf{Y}(\psi(\gamma(\mathbf{X}))$.

The success probability maximizing policy $\psi^{*}$ is therefore a solution to
\begin{align}
    \label{eq:success-proba}
    \underset{\psi\in \Delta^\mathcal{G}}{\max} \mathbb{P}\left(\sum_{u\in\mathcal{U}}\mathbf{Y}_u(\psi) \in \mathcal{S}\right) =  \underset{\psi\in \Delta^\mathcal{G}}{\max} \EE \left[\mathbb{I}_{\mathcal{S}} \left(\sum_{u\in\mathcal{U}}\mathbf{Y}_u(\psi)\right) \right], 
\end{align}

\vspace{-5pt}

where $\mathbb{I}_{\mathcal{S}}$ is the indicator function of the success set $\mathcal{S}$. 

\paragraph{Example} Our problem is defined with respect to any convex success region $\mathcal{S}\subset\mathcal{Y}$. In practice, we will consider success regions relative to a fixed $\mathbf{y}_0 = (y^v_0, y^c_0)$, of the form
$$
\mathcal{S}_{\mathbf{y}_0} = \{(y^v, y^c) \in \mathcal{Y} \ \text{s.t.} \ y^v > y^v_0 \ \text{and} \ y^c \leq y^c_0 \},
$$
where $\mathbf{y}_0$ should be interpreted as a \textit{reference outcome}, for example the outcome we observe if we assign the reference policy to every user $\sum_{u\in\mathcal{U}}\mathbf{Y}_u(\pi_0)$. The success region $\mathcal{S}_{\mathbf{y}_0}$ corresponds to all outcomes with an \textit{increased value and decreased cost} with respect to the reference value and cost $\mathbf{y}_0$.

In Figure~\ref{fig:sketch_spm}, $\mathcal{S}_{\mathbf{y}_0}$ is displayed in green and its complementary $\bar{\mathcal{S}}_{\mathbf{y}_0}$ in red. We represent the distributions of the outcome $\mathbf{Y}$ for the respective allocations output by (i) the possible solution to~\eqref{eq:success-proba} (maximization of the success probability) in orange and (ii) the possible solution to~\ref{eq:knapsack-grouped-expectation} (maximization of $\mathbb{E}[Y^v]$ under the condition $\mathbb{E}[Y^c] \leq y^c_0$) in blue. The orange outcome has a very high probability to be in $\mathcal{S}_{\mathbf{y}_0}$, even if it generates a bit less value on average than the blue one, which presents a high risk of being outside of the success region (for example by breaking the cost constraint). 

\section{The \texttt{SuccessProbaMax} algorithm}\label{sec:success-proba-max}

In this section, we present solutions for the problems \eqref{eq:knapsack-general-expectation} and~\eqref{eq:knapsack-grouped-expectation}. We will focus on the bucket-level versions of these problems, and therefore assume given a fixed partitioning $\gamma:\mathcal{X} \rightarrow \mathcal{G}=\{1, \dots, M\}$ of the feature space. This function could have been given by an expert or learned by machine learning algorithm, but it is not the focus of this work.

\subsection{Gaussian parametrization of the problem}\label{ssec:gaussian-param}
In this subsection, we introduce a novel method to solve the success probability maximization problem. This optimization problem, stated in \eqref{eq:success-proba}, presents several non-trivial difficulties:  (a) the indicator function which expectancy we are maximizing is not continuous on $\Delta^\mathcal{G}$ and (b) the criteria we wish to maximize is non-concave.
 We use 
\begin{align*}
    \mathbf{Y}_{g,k} 
    &= \sum_{u\in\mathcal{U}}\mathbf{Y}_u(\pi_{k})\mathbb{I}\left(\gamma(\mathbf{X}_u) = g\right),
\end{align*}
as a compact notation for the total expected outcome from users in bucket $g$, had they been allocated to policy $\pi_k$. 
Assuming that the buckets in $\mathcal{G}$ are approximately balanced in size (each containing $\approx N/M$ data points), we observe around $N / (M K)$ i.i.d. realizations to estimate each $\mathbf{Y}_{g,k}$. Let $\bmu_{k,g}$ and $\bSigma_{k,g}$ be the mean and covariance matrix of the potential outcomes which contain value and cost.

For any soft allocation $\psi: \mathcal{G} \rightarrow \Delta$ \textendash which maps all buckets in $\mathcal{G}$ to a stochastic mix of policies in $\Pi$\textendash the total expected outcome under allocation $\psi$ 
\begin{equation*}
\mathbf{Y}(\psi) = \sum_k \sum_g \psi(g, k) \mathbf{Y}_{g,k}, \quad \sum_k \psi(g, k) = 1.
\end{equation*}
The distributions $\psi(g, k) \mathbf{Y}_{g,k}$ are independent, therefore, we can use the Lyapunov central limit theorem and approximate the total expected outcome by a Gaussian distribution
\begin{equation*}
\mathbf{Y}(\psi) \sim \mathcal{N}\left(\bmu(\psi),
   \bSigma(\psi)
   \right),
\end{equation*}
where $\bmu(\psi) = \sum_{k} \sum_{g} \psi(g, k) \bmu_{g,k}$ and $ \bSigma(\psi) = \text{Var}\left[\sum_{k} \sum_{g} \psi(g, k) \textbf{Y}_{g,k} \right]$. Depending on assumptions, $\bSigma(\psi)$ can be a linear or quadratic function of $\psi(g, k)$ due to different sources of randomness which lead to different variances. We assume that $ \bSigma(\psi) = \sum_{k} \sum_{g} \psi(g, k) \bSigma_{g,k}$ (more details in Supplementary).

\subsection{Parameters estimation}

When we do not have a direct access to parameters $\bmu_{g, k}$ and $\bSigma_{g, k}$, we need to estimate them. In practice, the parameters are estimated on a \textbf{randomized control trial (RCT) dataset} $\mathcal{D} = \{(\mathbf{x}_u, \mathbf{y}_u)\}_{u\in\mathcal{U}}$ \textendash realization of the collection $\{(\mathbf{X}_u, \mathbf{Y}_u)\}_{u\in\mathcal{U}}$ introduced in the last section.
    More precisely, $(\mathbf{x}_u, \mathbf{y}_u) = \left(\mathbf{x}_u, \mathbf{y}_u(\pi_{k_u})\right)$ are i.i.d. realizations of $\left(\mathbf{X}, \mathbf{Y}(\pi_{k_u})\right)$, where $\{k_u\}_{u\in\mathcal{U}}$ are i.i.d. realization of a uniform categorical variable on $\{0, \dots, K-1\}$.
    For $k\in\llbracket 0, K-1\rrbracket$ and $g\in\mathcal{G}$, we will refer to the restrictions of $\mathcal{D}$ to points $u\in\mathcal{U}$ for which $\gamma(\mathbf{x}_u)=g$ and $k_u = k$ as $\mathcal{D}_{g,k}$. 

To estimate parameters, we choose
\textbf{mean and variance estimation methods} (e.g. bootstrapping~\citet{10.1214/aos/1176344552}) which take as input a dataset $\mathcal{D}_{g,k}$ containing realizations of $\mathbf{Y}$ for a given bucket $g$ and policy $k$ and return respectively its mean  $\{\hat{\bmu}_{g, k}\}$ and variances $\{ \hat{\bSigma}_{g, k} \}$.

\subsection{Gradient computation}
\label{sec:gradient_computation}

In the following, for $\psi\in\Delta^\mathcal{G}$, we will denote for clarity purposes $\mathcal{C}(\psi) = \mathbb{P}\left(\sum_{u\in\mathcal{U}}\mathbf{Y}_u(\psi)\in\mathcal{S}\right) = \EE \left[\mathbb{I}_{\mathcal{S}} \left(\sum_{u\in\mathcal{U}}\mathbf{Y}_u(\psi)\right) \right]$ the criterion we wish to optimize. The indicator function is discontinuous on the border of $\mathcal{S}$.  It prevents us from directly using a stochastic gradient method~\citep{shapiro2021lectures}. 
The next lemma \footnote{The proof of Lemma~\ref{lemma:reinforce} uses classical arguments from the policy learning literature~\citep{williams1992simple,sutton2018reinforcement} and further relies on the chain rule with a few relations for multivariate Gaussian variables. We defer the proof to the Supplementary. }
provides an explicit expression for the gradient of the criteria.

\begin{lemma}
\label{lemma:reinforce}
The gradient of $\mathcal{C}$  at $\psi$
satisfies
\begin{multline*}
       [\nabla \mathcal{C}(\psi)]_{g,k} =
       \EE \Big[\mathbb{I}_{\mathcal{S}}(\mathbf{Y}) \Big( (\mathbf{Y}-\bmu(\psi))^{\T}\bSigma(\psi)^{-1}\cdot \bmu_{g,k}\\
       -\frac{1}{2}\big(\bSigma(\psi)-(\mathbf{Y}-\bmu(\psi))(\mathbf{Y}-\bmu(\psi))^{\T}\big)
       \cdot
       \bSigma(\psi)^{-1} \bSigma_{g,k} \bSigma(\psi)^{-1}
        \Big)\Big]. 
\end{multline*}
\end{lemma}

\subsection{Optimization}
Here, we present our optimization algorithm \texttt{SuccessProbaMax} (Algorithm~\ref{alg:success-proba-gradesc}) to solve \eqref{eq:success-proba} which takes as input
\begin{itemize}
\item[\textbf{ (a)}] \textbf{success region } $\mathcal{S} \subset \mathcal{Y}$ to define a criteria $\mathcal{C}$ introduced in \eqref{eq:success-proba}. We typically consider success regions relative to a reference outcome $(y_0^v, y_0^c)$: it might be defined as all outcomes corresponding to increased value and
decreased cost with respect to the reference value and cost;
\item[\textbf{ (b)}] 
   \textbf{estimated mean and variance values} $\{\hat{\bmu}_{g, k}\}$ and $\{ \hat{\bSigma}_{g, k} \}$ for all pairwise couples of groups $g$ and candidate policies $k$. There is a particular case when the exact values of mean and variances are known and do not require estimation; 
\item[\textbf{ (c)}]
    \textbf{some hyperparameters} such as an initial policy allocation function $\psi_0\in\Delta^\mathcal{G}$, number of steps $n_{st}$ and learning rate~$\eta$.
 \end{itemize}   
\begin{algorithm}
\caption{\texttt{SuccessProbaMax}}
\label{alg:success-proba-gradesc}
\textbf{Input}: 
 $\mathcal{S}$, $\{\hat{\bmu}_{g, k}\}$, $\{ \hat{\bSigma}_{g, k} \}$, $\psi_0$, $n_{\text{st}} > 0$, $\eta > 0$\\
$\psi\gets\psi_0$\\
\For{$t=0$ to $n_{\text{st}}$}{
$\hat \bmu \gets \sum_{k, g}\psi(g,k) \hat{\bmu}_{g,k}$,
$\hat \bSigma \gets \sum_{k, g} \psi(g,k) \hat{\bSigma}_{g,k}$\\
$\nabla \gets \hat{\nabla} \mathcal{C}(\psi)$ \\
$\psi \gets \psi + \eta\nabla$ \\
Project $\psi$ onto $\Delta^M$
}
\textbf{Return} $\psi$
\end{algorithm}

\vspace{-5pt}

The algorithm performs a gradient ascent $\nabla \gets \hat{\nabla} \mathcal{C}(\psi)$ which can be computed using the formula from Lemma~\ref{lemma:reinforce} and a numerical integration method for computing the expectation $\EE_\psi$, e.g. a Monte-Carlo approach. 
The updated gradient is, then, projected onto the space of metapolicies $\Delta^\mathcal{G}$ to produce a solution candidate for~\eqref{eq:success-proba} using a method from~\citep{duchi2008efficient}. We provide several possible improvements of Algorithm \ref{alg:success-proba-gradesc} in Supplementary.
 
\paragraph{Remark}
As the computation of the gradient through the closed-form expression requires a matrix inversion, it is not always the best option computationally. 

This is the case for the  success region proposed in subsection~\ref{ssec:proba-max-problem}, for which we observe that 
the criterion rewrites
\begin{align*}
    \EE \left[\mathbb{I}_{\mathcal{S}} \left(\sum_{u\in\mathcal{U}} \mathbf{Y}_u(\psi) \right) \right] = {\rm{cdf}}_{Y^c}(y^c_0) - {\rm{cdf}}_\mathbf{Y}(\mathbf{y}_0), 
\end{align*}
where ${\rm{cdf}}_{Y_c}(y^c_0)$ is a (univariate) c.d.f. of $Y_c$ in $y^c_0$ and ${\rm{cdf}}_\mathbf{Y}(\mathbf{y}_0)$ is a (bivariate) c.d.f. of $\mathbf{Y}$ in $\mathbf{y}_0$.
(see appendix for the definition of bivariate c.d.f.).
To speed up the algorithm,  we rely on an approximation of the bivariate c.d.f. based on the error function~\citep{tsay2011simple} to estimate ${\rm{cdf}}_\mathbf{Y}(\mathbf{y}_0)$, and then implement it in JAX – this way we can directly use the automatic differentiation in JAX to numerically approximate the gradient of ${\rm{cdf}}_{Y^c}(y^c_0) - {\rm{cdf}}_\mathbf{Y}(\mathbf{y}_0)$.

\section{Experimental Results}
\label{sec:experiments}

For all experiments below we use JAX framework~\citep{jax2018github} for the numerical estimation of the criterion's gradient, utilizing automatic differentiation within JAX instead of explicitly calculating the gradient and integrating it over an outcome.
Hyperparameters used for the methods are provided in Supplementary material and source code\footnote{https://github.com/criteo-research/success-proba-max} is published to reproduce all the empirical results.

\subsection{Datasets}
Besides the synthetic setups, which will be described below, we test algorithm on two large-scale, real world datasets.
\begin{itemize}
    \item \textbf{CRITEO-UPLIFT v2} \cite{diemert2021large} is provided by the AdTech company Criteo. Data contains 13.9 million samples which are collected from several incremental A/B tests. It includes 12 features, 1 binary treatment and 2 binary outcome labels ("visit" and "conversion"). Following \cite{zhou2023direct}, we use "visit" label as proxy of the cost and "conversion" as the value. 
    For the buckets, we used quantile bins of the "f0" feature. Finally, we randomly partitioned dataset into two equal parts for train and test. 
    Preprocessing details are in Supplementary.

    \item \textbf{Private dataset} is constructed from a large-scale real-time bidding RCT. 
    One feature was chosen based on an expert knowledge, buckets were created then as quantile-based projections of the feature.
    Dataset is aggregated over 70 days and consists of 9 buckets, 3 bidding policies (with reference) and 100 bootstraps of values and associated costs for each pair (bucket, policy).
    
    Remaining details along with aggregated datasets for one- and two-dimensional outcome cases are available in Supplementary material.
\end{itemize}

\subsection{One-dimensional outcome}
Here we assume an outcome $\mathcal{Y} \in \mathbb{R}$. Problem is parameterized by a difficulty level $r$ so that $\mathcal{S}=\{(r,+\infty)\}$. 
We present here results for synthetic data. Private data results are in Supplementary.

\paragraph{Baselines} \texttt{SuccessProbaMax} is compared to several baselines searching for the optimal policy allocation:
\begin{itemize}
    \item \texttt{Bruteforce}$(\{\bmu_{g,k}\}, \{\bSigma_{g,k}\}, \mathcal{S})$ method that compares all possible \textit{hard} allocations and for a given difficulty level returns allocation that maximises criterion; 
    \item \texttt{Greedy1D}$(\{\bmu_{g,k}\})$ algorithm that returns the policy with the maximum mean value per bucket.
\end{itemize}

\subsubsection{Synthetic data generation}
We generate Gaussian distributions for two cases: (i) "large variance" and 
(ii) "small variance", the same setup but the relative difference between the variances is much smaller – we expect the latter problem be harder than the former for the algorithms that take into account the variance. 
See Table~\ref{table:1d_parameters} for precise parameters of distributions (data construction details and illustration of policy distributions per bucket are provided in Supplementary). 

\begin{table}[!ht]
\caption{Gaussian distribution parameters for synthetic data generation with three buckets ($M=3$), three policies ($K = 3$) and one outcome ($Y \in \mathbb{R}$).}
\label{table:1d_parameters}
\centering
\small{
\resizebox{0.7\columnwidth}{!}{%
\begin{tabular}{|c|l|l|}
\hline
Example        & \multicolumn{1}{c|}{$[\mu_{g,k}]$}                                                                               & \multicolumn{1}{c|}{$[\Sigma_{g,k}]$}                                                                                          \\ \hline
Large variance & $\begin{bmatrix} 2 & 1.9 & 0 \\ 2 & 1 & 0 \\ 2 & 1 & 0 
\end{bmatrix}$    & $\begin{bmatrix} 9 & 1 & 9   \\ 9 & 1 & 9 \\ 1 & 1 & 1 \end{bmatrix}$ \\ \hline
Small variance & $\begin{bmatrix} 2 & 1.9 & 0 \\ 2 & 1 & 0 \\ 2 & 1 & 0 
\end{bmatrix}$    & $ \begin{bmatrix} 9 & 1 & 9   \\ 9 & 1 & 9 \\ 1 & 1 & 1 \end{bmatrix} \cdot 0.01 $  \\ \hline
\end{tabular}
}}
\end{table}

\paragraph{Results}
We firstly fix $\mu_{g, k}$, $\Sigma_{g, k}$ and use them directly in the algorithm to avoid a source of randomness arising from parameters estimation, we provide the results below. Then we generate normal distributions with parameters $\mu_{g, k}$, $\Sigma_{g, k}$ and use \textit{estimations} $\hat{\mu}_{g, k}$, $\hat{\Sigma}_{g, k}$ in the algorithm – corresponding results are presented in Supplementary. 

On Fig.~\ref{fig:1d_synth_res} (left) we show how our method performs for easier problem with large variance with varying difficulty level $r$. \texttt{SuccessProbaMax} starts from uniform allocation $\psi_0^{unif}$ and performs same as \texttt{Bruteforce}. The key reasons why our algorithm beats \texttt{Greedy1D} is that we i) directly optimize metric of interest and that we ii) effectively incorporate variance into optimization, while \texttt{Greedy1D} only operates with means.

Fig.~\ref{fig:1d_synth_res} (right) shows results for the small variance case. Firstly, note that the gain over \texttt{Greedy1D} (in the region $r \in [5,5.5]$) is drastically smaller than in the previous case. Then, at difficulty level $r > 6$ the performance of our algorithm drops down. This is because criterion value $\mathcal{C}(\psi_0^{unif})$ becomes 0 and the gradient is not updated. To overcome the problem, we can either "warm-start" from the baseline policy (e.g. from \texttt{Greedy1D} one) or to explore, by estimating the criterion for several random initial allocations. 

\begin{figure}[h!]
  \begin{center}
    \includegraphics[scale=0.275]{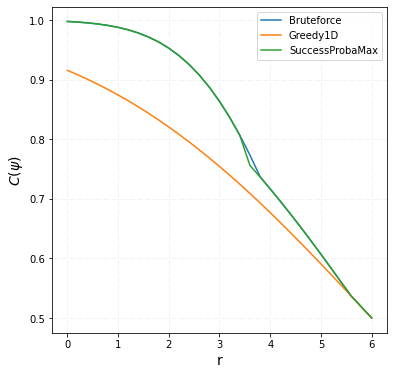}
    \hspace{-0.25cm}
    \includegraphics[scale=0.275]{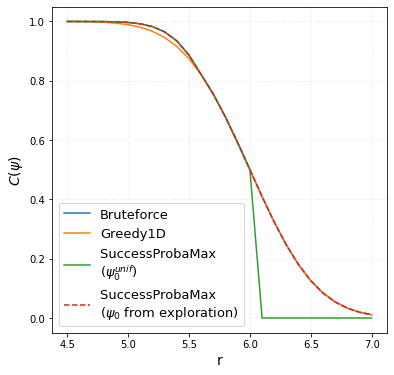}
  \end{center}
  \caption{Results for different difficulty level $r$ on synthetic setup with one-dimensional outcome for large (left) and small (right) variance cases.}
  \label{fig:1d_synth_res}
\end{figure}

\subsection{Two-dimensional outcome}

In this case we consider an outcome $\mathcal{Y} \in \mathbb{R}^2$, so $\mathbf{Y} = (Y^v, Y^c)$. Problem is parameterized by two-dimensional difficulty levels $r$ so that $\mathcal{S}=\{(r_v,+\infty), (-\infty, r_c]\}$. 

\paragraph{Baselines} In addition to \texttt{Bruteforce}, \texttt{SuccessProbaMax} is compared with two other baselines that search for the optimal policy allocation:
\begin{itemize}
    \item \texttt{LinProg}$(\{\bmu_{g,k}\}, r_c)$ algorithm (linear programming) that solves the fractional knapsack problem and returns a policy (soft allocation) with the maximum mean value per bucket;
    \item \texttt{MixedInt}$(\{\bmu_{g,k}\}, r_c)$ algorithm (mixed-integer linear programming) that solves the 0/1 knapsack problem and returns a policy (hard allocation) with the maximum mean value per bucket.
\end{itemize}

\subsubsection{Synthetic data generation}

We generate bivariate Gaussian distributions for cases (i) and (ii), see Table~\ref{table:2d_parameters} for precise parameters of distributions (an illustration of policy distributions is provided in Supplementary). 

\begin{table}[h!]
\caption{Bivariate Gaussian distribution parameters for synthetic data generation with one bucket ($M=1$), two policies ($K = 2$) and two-dimensional outcome ($\mathbf{Y} = (Y^v, Y^c) \in \mathbb{R}^2$).}
\label{table:2d_parameters}
\centering
\small{
\resizebox{0.95\columnwidth}{!}{%
\begin{tabular}{|c|c|c|c|c|c|}
\hline
Example                           & $[\mu^v_{g,k}]$                      & $[\Sigma^v_{g,k}]$ & $[\mu^c_{g,k}]$                        & $[\Sigma^c_{g,k}]$ & $\rho$ \\ \hline
\multicolumn{1}{|c|}{$\mu_2^c$ and $\Sigma_1^c$ larger  } & \multicolumn{1}{|c|}{$[2, 1]$} & $[9, 1]$    & \multicolumn{1}{|c|}{$[1, 1.5]$} & $[4, 1]$ &  0.5    \\ \hline
\multicolumn{1}{|c|}{$\mu_2^c$ and $\Sigma_1^c$ smaller  } & \multicolumn{1}{|c|}{$[2, 1]$} & $[9, 1]$    & \multicolumn{1}{|c|}{$[1, 0.5]$} & $[1, 1]$ &  0.5    \\ \hline
\end{tabular}
}}
\end{table}

We firstly fix $\mu^v_{g, k}, \Sigma^v_{g, k}$ and $\mu^c_{g, k}, \Sigma^c_{g, k}$, and use them directly in the algorithm to avoid a source of randomness arising from parameters estimation (results for the case with parameters estimation are presented in Supplementary). 

\begin{figure}[h]
  \begin{center}
    \includegraphics[scale=0.27]{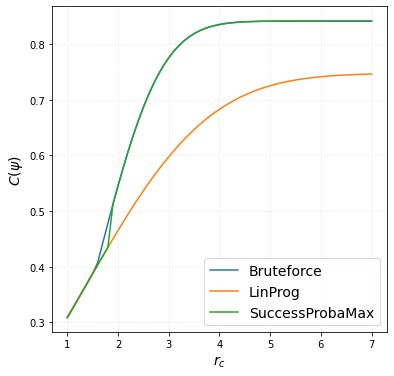}
    \hspace{-0.25cm}
    \includegraphics[scale=0.27]{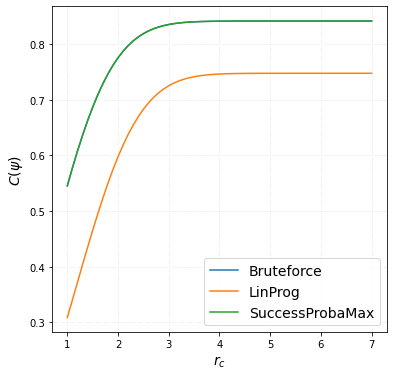}
  \end{center}
  \caption{Result for different $r_c$ with fixed $r_v = 0$ on synthetic setup with two-dimensional outcome for cases (i) (left) and (ii) (right).}
  \label{fig:2d_synth_res}
\end{figure}

\paragraph{Two-dimensional outcome: results.}
For the experiment, we fix $r_v = 0$ and vary $r_c$ only. Fig.~\ref{fig:2d_synth_res} shows that for  both cases, \texttt{SuccessProbaMax} started from uniform allocation $\psi_0^{unif}$ reaches the same performance as \texttt{Bruteforce}. 

\subsubsection{Private dataset}
In our first experiment, we fix $r_c = 0$ and vary $r_v$ only - so we check if we can increase total value while having same total cost as for reference. We then repeat computations, but now we fix $r_v = 0$ and vary $r_c$ only - in this case we wonder how often we can reach at least total value of the reference policy while changing total cost (this case is described in Supplementary).

\paragraph{Results}
Fig.~\ref{fig:priv_2d_res_v} describes results on the private dataset with two-dimensional outcome for the range of Gain $r_v$ while $r_c = 0$ for train (left) and test (right) splits. Our algorithm, initialized with $\psi_0$ from exploration, reach the Gain of 0.01 in value (1\% over the reference) with probability 0.7 for train and 0.4 for test, while for \texttt{MixedInt} respective probabilities are 0.35 and 0.1. Note that \texttt{Bruteforce} might not be the best here as some soft allocation may outperform hard ones for particular $(r_v, r_c)$.
\begin{figure}[h!]
  \begin{center}
    \includegraphics[scale=0.27]{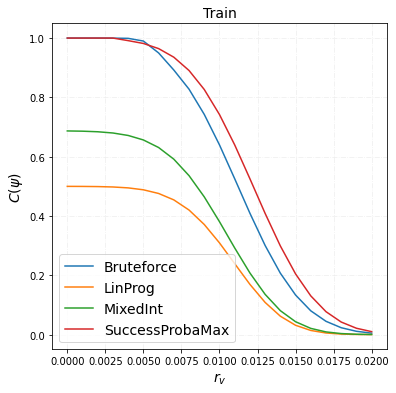}
    \hspace{-0.25cm}
    \includegraphics[scale=0.27]{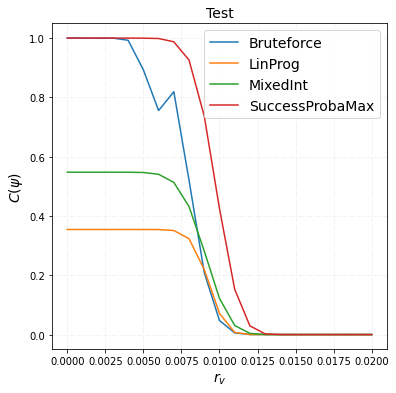}
  \end{center}
  \caption{Results for different Gain $r_v$ while $r_c = 0$ on private dataset with one-dimensional outcome for train (left) and test (right) splits.}
  \label{fig:priv_2d_res_v}
\end{figure}

\subsubsection{CRITEO-UPLIFT v2}
\begin{figure}[h!]
  \begin{center}
    \includegraphics[scale=0.30]{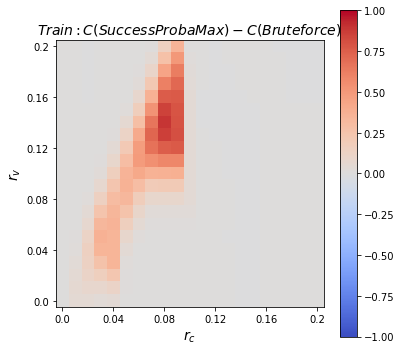}
    \hspace{-0.25cm}
    \includegraphics[scale=0.30]{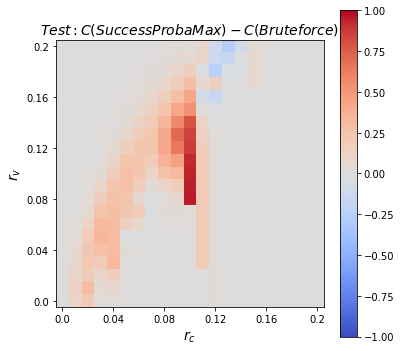}
  \end{center}
  \caption{Results for different $r_v$ and $r_c$ on CRITEO-UPLIFT v2 with two-dimensional outcome for train (left) and test (right) splits.}
  \label{fig:cu2_2d_res}
\end{figure}

Data contains 2 policies including reference ("control"), so value can be increased only by increasing the cost. Thus, now we vary both $r_v$ and $r_c$ from 0 to 0.2, and a trade-off between value and cost is expected.
\paragraph{Results}
Fig.~\ref{fig:cu2_2d_res} depicts differences in $\mathcal{C}(\psi)$ between our algorithm and best baseline \texttt{Bruteforce} (absolute values are provided in Supplementary). Firstly, there is indeed a trade-off - for increasing cost by $x\%$ value increases by roughly $2x\%$. In addition, our algorithm reach higher $\mathcal{C}(\psi)$ in several regions (e.g. where $r_c \in [0.03, 0.04]$ and $r_v \in [0.04,0.08]$ or where $r_c \in [0.08, 0.1]$ and $r_v \in [0.1,0.16]$). 

The results correspond to the sketch provided in Fig.~\ref{fig:sketch_spm}. It represents the distributions of the outcome $\mathbf{Y}$ for the respective allocations output by (i) \texttt{SuccessProbaMax} in orange and (ii) \texttt{Greedy} in blue. \texttt{SuccessProbaMax} outputs a solution for which the outcome has a very high probability to be in $\mathcal{S}_{\mathbf{y}_0}$ with $\mathbf{y}_0 = (y_0^v, y_0^c) = (r_v, r_c)$, even if generates a bit less value on average than \texttt{Greedy}, which presents a high risk of being outside of the success region (for example by breaking the cost constraint).

\section{Related work}\label{sec:related-work}
Some recent papers address multiple treatment allocation problem under budget constraints from different perspectives. 
The standard two-stage method firstly estimates treatment effects to predict value and cost for each user, then solves a knapsack problem~\citep{ai2022lbcf,albert2022commerce,tu2021personalized,zhao2019unified}.
Nevertheless, the goals of two-stage approaches and real-world scenarios do not perfectly align. \citet{yan2023end} proposes a two-stage method with an addition regularizer to the knapsack problem loss to address a business goal. However, the regularizer requires a mathematically well-defined function (such as expected outcome metric) and its gradients estimation. 
 
Applying the decision-focused framework for marketing problems under budget constraints, \citet{du2019improve} propose a rank method by comparing learned ratios between values and costs for the aggregated targeted treatment effect to improve user retention problem. However, \citet{zhou2023direct} show that the suggested loss function cannot converge to a stable extreme point in theory and improve the framework.
Authors limit the treatments to different levels of one treatment, e.g. different levels of discount of some products. Further, they develop an algorithm equivalent to the Lagrange dual method ('greedy' approach) but based on learning to rank decision factors for multiple choice knapsack problem solutions. In our current context, our focus is solely on the top-ranked action, rather than the complete ranking itself. Moreover, as we discussed earlier, the knapsack formulation remains a proxy to our problem, so finding efficiently the best decision factors is still not equivalent to finding the best solution to the final business goal. 

The closest to our work, \citet{tu2021personalized} suggest to reformulate the treatment allocation problem as a stochastic optimization task assuming normally distributed outcomes of bucket-level objective and constraints, however, the final problem remains in the knapsack form.

\section{Conclusion and future works}
\label{sec:conclusion}
We suggested a new formulation of the policy allocation problem that is better adapted to some downstream tasks when the success region is clearly identified. Compared to greedy approaches, our algorithm directly optimizes metric of interest and effectively utilizes variance in the optimization, while greedy ones only operate with means. 
Moreover, the proposed method can be efficiently applied to improve the given baseline policy.

Further works include a theoretical analysis of the algorithm. In particular, how it behaves numerically when the dimension of the outcome increases.
Also, it is important to understand the relationship between the means and variances of potential outcomes that makes the proposed method  outperform the greedy approaches. 
In addition to several suggested improvements of Algorithm~\ref{alg:success-proba-gradesc}, promising direction should be to couple the choice of user partitioning and the policy allocation problem into one master problem. 
Last, given that outcomes on different user segments may correlate, adapting the framework for Bayesian learning seems a pragmatic avenue for further research. 

\section{Acknowledgments}

We would like to thank David Rohde and Eustache Diemert for their feedback and ideas during the project. 


\appendix
\section{Discussion on different sources of variance}

There are different sources of randomness: 
\begin{itemize}
    \item \textit{estimation variability}: how close is the estimator $\hat{\mathbf{Y}}_{g,k}$ to the true value $\Ygk$ (when we do not have a direct access to $\Ygk$);
    \item \textit{allocation variability}: when $\psi$ is a soft allocation, the allocation of a user $u\in g$ to policy $\in\mathcal{K}$ can be described by the categorical random variable $P^\psi(g)\in \mathcal{K}$ such that $\mathbb{P} \left( P^\psi(g) = k \right) = \psi(g, k)$. We call $\mathcal{P}$ the filtration containing all the randomness from the variables $P^\psi(g)$ for all $g$.
    \item \textit{system stochasticity}: for a fixed $g$ and $k$, the quantity $\Ygk$ is itself a random variable which contains the randomness from the behaviour of users in $g$. We call $\mathcal{Y}$ the filtration containing all randomness from the variables $\Ygk$ for all $g, k$.
\end{itemize}

In our work, we do not consider the problem of estimation variability, but focus on the derivation of the variance of $\mathbf{Y}$ under a given soft allocation $\psi$.
This variance $ \Var[\mathbf{Y}(\psi)]$ can be decomposed according to the variance with respect to the two sources of stochasticity respectively contained in $\mathcal{Y}$ and $\mathcal{P}$.
\begin{align}
\label{eq:vardecompP}
    \Var[\mathbf{Y}(\psi)] &= \mathbb{E}_{\mathcal{P}} \left[ \Var_{\mathcal{Y}} \left[\bY | \mathcal{P} \right] \right] + \Var_{\mathcal{P}} \left[ \mathbb{E}_{\mathcal{Y}} \left[ \bY | \mathcal{P}\right] \right] \\
\label{eq:vardecompY}
    &= \mathbb{E}_{\mathcal{Y}} \left[ \Var_{\mathcal{P}} \left[\bY | \mathcal{Y} \right] \right] + \Var_{\mathcal{Y}} \left[ \mathbb{E}_{\mathcal{P}} \left[ \bY | \mathcal{Y}\right] \right]
\end{align}

Depending on assumptions, we can have different approximations of $\Var[\mathbf{Y}(\psi)]$ as a linear or quadratic function of $\psi$. For example, consider the decomposition of the variance with respect to allocation randomness in \eqref{eq:vardecompP}:
\begin{align*}
\mathbb{E}_{\mathcal{P}} \left[ \Var_{\mathcal{Y}} \left[\bY | \mathcal{P} \right] \right] &= \sum_k \sum_g \psi(g, k) \bSigma_{g, k}, \\
\Var_{\mathcal{P}} \left[ \mathbb{E}_{\mathcal{Y}} \left[ \bY  | \mathcal{P}\right] \right] &= \sum_g \psi(g, k) \bigl( \bmu_{g, k}  - \sum_k \psi(g, k) \bmu_{g, k}\bigr)^2.
\end{align*}
If we assume that there is minimal allocation variability, then $\Var_{\mathcal{P}} \left[ \mathbb{E}_{\mathcal{Y}} \left[ \bY  | \mathcal{P}\right] \right] << \mathbb{E}_{\mathcal{P}} \left[ \Var_{\mathcal{Y}} \left[\bY | \mathcal{P} \right] \right]$ and the variance of $\mathbf{Y}$ depends linearly on $\psi$ since
\begin{equation*}
     \Var[\mathbf{Y}(\psi)] \approx 
\mathbb{E}_{\mathcal{P}} \left[ \Var_{\mathcal{Y}} \left[\bY | \mathcal{P} \right] \right] 
= \sum_k \sum_g \psi(g, k) \bSigma_{g, k}.
\end{equation*}
In practice, we assume that the allocation variability is small enough so that this variance approximation holds, i.e. we are capable of assigning a given ratio of the population to a given policy. Indeed, we observe that for all $g,k$ there are enough users in $g$ such that we have an (approximately) fixed proportion $\psi(g,k)$ of users from $g$ which are allocated to policy $k$.

\section{Gradient derivation (proof of Lemma 1)}

\begin{enumerate}
\item We have 
$    \nabla \mathcal{C}(\psi) =\EE \left(\mathbb{I}_{\mathcal{S}}(\bY) \nabla_\psi\left(\ln \ell(\psi, \bY)\right) \right).
$
\begin{proof}
   \begin{align*}
\nabla \mathcal{C}(\psi) &= \nabla \int_{\mathcal{S}} \mathbb{I}_{\mathcal{S}}(\by) \ell(\psi, \by) \dd \by \\
&= \int_{\mathcal{S}} \mathbb{I}_{\mathcal{S}}(\by) \nabla\ell(\psi, \by) 
\dd \by \\
&= \int_{\mathcal{S}} \mathbb{I}_{\mathcal{S}}(\by) \frac{\nabla_\psi\ell(\psi, \by)}{\ell(\psi, \by)}\ell(\psi, \by)\dd \by\\
&=\int_{\mathcal{S}} \mathbb{I}_{\mathcal{S}}(\by) \nabla_\psi\left(\ln \ell(\psi, \by)\right) \ell(\psi, \by)\dd \by \\
&= 
\EE \left(\mathbb{I}_{\mathcal{S}}(\by) \nabla_{\psi}\left(\ln \ell(\psi, \by)\right) \right)
    \end{align*}
\end{proof}
We need the derivatives of a Gaussian log-likelihood of with respect to  its parameters. 

\item
Let $\rm{p}[\bmu,\bSigma]$ be the probability density function of $\mathcal{N}(\bmu,\bSigma)$, then 
\begin{align*}
     \nabla_{\bSigma^{-1}}\left(\ln \rm{p}[\bmu,\bSigma]\by)\right) &=\frac{1}{2}\bSigma-\frac{1}{2}(\by-\bmu)(\by-\mu)^{\T}\ , \\
         \nabla_{\bmu}\left(\ln \rm{p}[\bmu,\bSigma](\by)\right) &=(\by-\bmu)^{\T}\bSigma^{-1}.
\end{align*}
\begin{proof}
The proof is based on the rules of derivation by a vector and inversed matrix. 
\end{proof}
\item
If $f:t\to A(t)$ is an application from $\RR$ to the set of non singular matrix of dimension $d$, then 
 the derivative of $g:t\to A^{-1}(t)$ is $-g f' g$. 
 
\item Let $\ell(\psi, \by) = \rm{p}[\bmu(\psi),\bSigma(\psi)](\by)$, then 
\begin{multline*}
       \left( \nabla_{\psi}\left(\ln \ell(\psi, \by)\right)\right)_{g,k} = (\by-\bmu(\psi))^{\T}\bSigma(\psi)^{-1}\cdot \bmu_{g,k} \\
       -  \frac{1}{2}
       \left(\bSigma(\psi)-(\by-\bmu(\psi))(\by-\bmu(\psi))^T\right) \\ 
       \quad \cdot 
       \bSigma(\psi)^{-1} \bSigma_{g,k} \bSigma(\psi)^{-1}.
\end{multline*}
\begin{proof}
Using the fact that 
$\bmu(\psi)=\sum_{g,k}  \psi(g,k) \bmu_{g,k}$ and $\bSigma(\psi)=\sum_{g,k}  \psi(g,k) \bSigma_{g,k}$, it is clear that
\begin{align*}
    \frac{\partial\bmu(\psi)}{\partial \psi(g,k)}&=\bmu_{g,k},\\
    \frac{\partial \bSigma(\psi)}{\partial \psi(g,k)}&= \bSigma_{g,k}.
\end{align*}
Let 
$\ell(\psi, \by) = \rm{p}[\bmu(\psi),\bSigma(\psi)](\by)$ and index $(g,k)\in M\times K$, then,  the chain rule and the previous steps lead to
    \begin{align*}
       \frac{ \partial\left(\ln \ell(\psi,\by)\right)}{\partial \psi(g,k)} 
       &= \frac{ \partial\left(\ln \rm{p}[\bmu(\psi),\bSigma(\psi)](\by)
       \right)}{\partial \psi(g,k)}\\
       &=\frac{\partial\left(\ln \rm{p}[\bmu(\psi),\bSigma(\psi)](\by)
       \right)}{\partial \bmu(\psi)}\frac{\partial\bmu(\psi)}{\partial \psi(g,k)} \\
       & \ \ +\frac{\partial\left(\ln \rm{p}[\bmu(\psi),\bSigma(\psi)](\by)
       \right)}{\partial \bSigma(\psi)^{-1}}\frac{\partial \bSigma(\psi)^{-1}}{\partial \psi(g,k)},
    \end{align*}
    where
    \begin{align*}
       \frac{\partial\left(\ln \rm{p}[\bmu(\psi),\bSigma(\psi)](\by)
       \right)}{\partial \bmu(\psi)}\frac{\partial\bmu(\psi)}{\partial \psi(g,k)} \\
       =(\by-\bmu(\psi))^T\bSigma(\psi)^{-1} \bmu_{g,k},
    \end{align*}
    \begin{align*}
       &\frac{\partial\left(\ln \rm{p}[
       \bmu(\psi),\bSigma(\psi)] (\by)
       \right)}{\partial \bSigma(\psi)^{-1}} \frac{\partial \bSigma(\psi)^{-1}}{\partial \psi(g,k)} = \\
       &= \frac12 \left(\bSigma(\psi)-(\by-\bmu(\psi))(\by-\bmu(\psi))^T\right) \\
       &\ \ \cdot
       \left(
       \bSigma(\psi)^{-1} \bSigma_{g,k} \bSigma(\psi)^{-1}
       \right).
    \end{align*}
\end{proof}
\item The gradient of index $(g,k)\in M\times K$ is the following
\begin{align*}
       [\nabla \mathcal{C}(\psi)]_{g,k} &=
       \mathcal{C}(\psi)\cdot \EE_{\psi}\bigl( (\by-\bmu(\psi))^{\T}\bSigma(\psi)^{-1} \cdot \bmu_{g,k} \\
       &-\frac{1}{2}\left(\bSigma(\psi)-(\by-\bmu(\psi))(\by-\bmu(\psi))^T\right)\\ 
       &\cdot 
       \left(
       \bSigma(\psi)^{-1} \bSigma_{g,k} \bSigma(\psi)^{-1}
       \right)|\by\in\mathcal{S}\bigr)
\end{align*}
\end{enumerate}
    
\noindent Let $M=1$ and $d=1$, we get 
\begin{align*}
      \frac{ \partial_k \mathcal{C}(\psi)}{\mathcal{C}(\psi)} &=
       \EE_{\psi}\Biggl[ \mathbb{I}_{\mathcal{S}}(y)\Biggl( \mu_{k}\frac{(y-\mu(\psi))}{\Sigma(\psi)} \\
       &-\frac{\Sigma_{k}}{2}\frac{\left(\Sigma(\psi)-(y-\mu(\psi))^2\right)}{\Sigma(\psi)^2} 
       \Biggr) \Biggr].
\end{align*}

\section{Improvements of \texttt{SuccessProbaMax}}
We identify several directions of how \texttt{SuccessProbaMax} can be improved. Firstly, gradient step may be accelerated, either by i) using second-order methods like Newton method, ii) by applying line search to adapt step size. Secondly, we observe that algorithm may be stuck in the "flat" regions, e.g. if the criterion value of the initial policy allocation equals 0 – this problem often appears in policy gradient methods in reinforcement learning ~\citep{schulman2015trust,levine2020offline}. Currently, we explore several random allocations to begin optimization from (akin to epsilon-greedy exploration in reinforcement learning) or "warm-start" from baseline policy, but there are more options to avoid this behaviour, e.g. forcing exploration by regularization.

\section{Example of alignment}

Here, we provide an example of a problem when \texttt{SuccessProbaMax} and \texttt{Greedy} give the same solution.

Consider two policies $\pi_0$ and $\pi_1$ and users from $\mathcal{U}$ with a population of size $N$. Let potential outcome $Y_u(\pi_k)$ of user $u$ follow Bernoulli distributions $\mathcal{B}(p_k)$, where $k \in \{0, 1\}$. Our goal is to maximize \textit{success} $\mathbb{P} \left( \sum_u Y_u (\pi_{k_u}) = r \right)$, i.e. the probability of getting exactly $r$ successes in $N$ independent Bernoulli trials with parameters $p_0$ or $p_1$ depending on which policy $\pi_{k_u} \in \{\pi_0, \pi_1\}$ is assigned to users $u \in \mathcal{U}$. 

If parameters $p_0$ and $p_1$ are not known, we need to estimate them from the data. Let $N_0$ users be assigned policy $\pi_0$ and $N_1 = N - N_0$ users be assigned policy $\pi_1$. For each policy, we observe $y(\pi_k) = \sum_{u = 1}^{N_k} y_u(\pi_k)$, a realization of $Y(\pi_k) \sim \mathrm{Binom}(N_k, p_k)$, where $y_u(\pi_k)$ are sampled from $\mathcal{B}(p_k)$. We estimate the Bernoulli probability for each policy as $\hat{p}_k = \frac{1}{N_k} y(\pi_k)$.
If we assume that there is no variance due to \textit{estimation variability}, i.e. $|\hat{p}_k - p_k| \approx 0$, the total variance of $Y$ after observing $y = y(\pi_0) + y(\pi_1)$ is due to \textit{system stochasticity} that consists of variance coming from the binomial distribution:
\begin{align*}
    \mathrm{Var}[Y] &= \mathrm{Var}[Y(\pi_0)] + \mathrm{Var}[Y(\pi_1)] \\
    &= N_0 \hat{p}_0 (1 - \hat{p}_0) + N_1 \hat{p}_1 (1 - \hat{p}_1)
\end{align*}
We notice that the variance of the total outcome $\mathrm{Var}[Y]$ is monotone with respect to $N_0$ and $N_1$. 

In this example, \texttt{SuccessProbaMax} will search for a trade-off between the policy with the minimum estimated Bernoulli variance 
$\arg_k\min \hat{p}_k (1 - \hat{p}_k)$ and the maximum Bernoulli mean $\arg_k\max \hat{p}_k$. Since $\arg_k\min \hat{p}_k (1 - \hat{p}_k) = \arg_k\max \hat{p}_k$, we obtain that, in this example, \texttt{SuccessProbaMax} returns the same solution as \texttt{Greedy}. 

\section{Datasets}
Here we describe two real datasets used in the paper.

\subsection{Private Dataset}
Data was created from a large-scale real-time bidding randomized control trial (RCT) over 70 days and consists of 3 labels. The main label is the \textit{value} - originally binary variable of some user action. For each value we collected an associated \textit{cost}. We used \textit{value} and \textit{cost} for \textit{two-dimensional} experiments. Along with this, we approximated a \textit{revenue} as a function of value and cost, which will be used for \textit{one-dimensional} experiments. Data consists of 3 randomly (respecting the RCT procedure) assigned policies, $\{\pi_0, \pi_1, \pi_2\}$, where $\pi_0$ is a reference bidding policy used in production, $\pi_1$ and $\pi_2$ are candidate bidding policies. In order to separate the user-level feature space, one feature was chosen based on an expert knowledge, 9 buckets were created then as quantile-based projections of the feature.

We aggregated labels by summarising them across the triplets (day, bucket, policy). Along with the sums, we computed 100 bootstraps of the aggregated value, cost and revenue, that will be used for the mean and (co-)variance estimations.

In order to make a balanced yet realistic train/test split, we summed labels for odd (train) and even (test) calendar days, hence we got both train and test data aggregated over 35 days.   

To maintain data confidentiality, we computed a relative difference of labels with respect to the reference policy – for each pair (bucket, policy) we subtracted a value of the reference policy from the original one and divided it by a total reference value (a sum over buckets), we did the same for the cost and revenue. Finally, we used resulted bootstraps to estimate $\hat{\bmu}_{g,k}$ and $\hat{\bSigma}_{g,k}$.

\subsection{CRITEO-UPLIFT v2}
Dataset is provided by the AdTech company Criteo. Data contains 13.9 million samples which are collected from several incremental A/B tests. It includes 12 features, 1 binary treatment and 2 binary outcome labels ("visit" and "conversion"). Following \cite{zhou2023direct}, we use "visit" label as proxy of the cost and "conversion" as the value. We randomly partitioned the dataset into two equal parts for train and test. For the buckets, we used quantile bins of the "f0" feature resulting in 8 buckets. 

\section{One-Dimensional outcome}
\subsection{Criterion}
Here we assume an outcome $\mathcal{Y} \in \mathbb{R}$. The problem is parameterized by a difficulty level $r$ so that $\mathcal{S}=\{(r,+\infty)\}$. 
Criterion then is defined as maximizing the following probability
\begin{align*}
    \mathbb{P}(Y(\psi) > r) = \EE \left[\mathbb{I}_{\mathcal{S}} \left(\sum_{u\in\mathcal{U}} Y_u(\psi) \right) \right] = 1 - {\rm{cdf}}_{Y(\psi)}(r), 
\end{align*}
where 
\begin{align*}
Y(\psi) & = \sum_{k} \sum_{g} \psi(g, k) Y_{g, k} \sim \mathcal{N}(\mu(\psi), \Sigma(\psi)), \\
\mu(\psi) & = \sum_{k} \sum_{g} \psi(g, k) \mu_{g, k}, \\
\Sigma(\psi) & = \sum_{k} \sum_{g} \psi(g, k) \Sigma_{g, k}, 
\end{align*}
and ${\rm{cdf}}_{Y(\psi)}(r)$ is a cumulative distribution function (c.d.f.) of $Y(\psi)$ in $r$.

Instead of computing the gradient as an integration over $Y(\psi)$ to obtain the expected value (see Lemma 1), we use an automatic differentiation in JAX to numerically approximate the gradient of $1 - {\rm{cdf}}_{Y(\psi)}(r)$.

\subsection{Baselines: \texttt{Bruteforce}}
This method compares all possible \textit{hard} allocations and for a given difficulty level returns allocation that maximises criterion, thus, resulting in the complexity $O(K^M)$, where $K$ is a number of policies and $M$ is a number of buckets – for large $K$ and $M$, Bruteforce is not an option because of its non-polynomial complexity.

\subsection{Synthetic setup}
We simulate a toy yet sufficient setup in order to illustrate typical situations in which our algorithm can make an advantage.

\begin{figure}[h!]
  \begin{center}
    \includegraphics[scale=0.3]{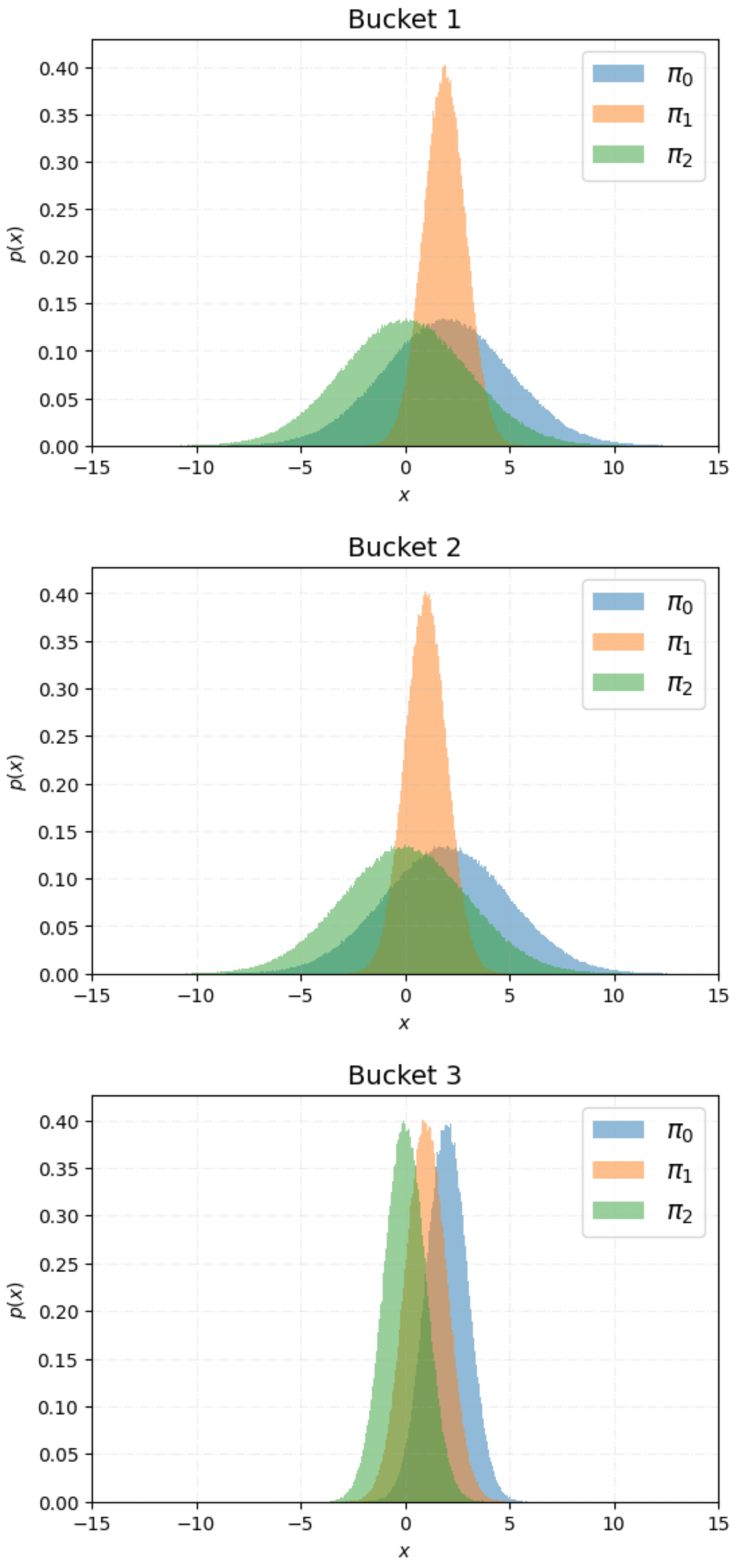}
  \end{center}
  \caption{Synthetic data distributions for large variance case, with parameters described in Table~\ref{table:1d_parameters}.}
  \label{fig:1d_large_var_distr}
\end{figure}

\begin{figure}[h!]
  \begin{center}
    \includegraphics[scale=0.3]{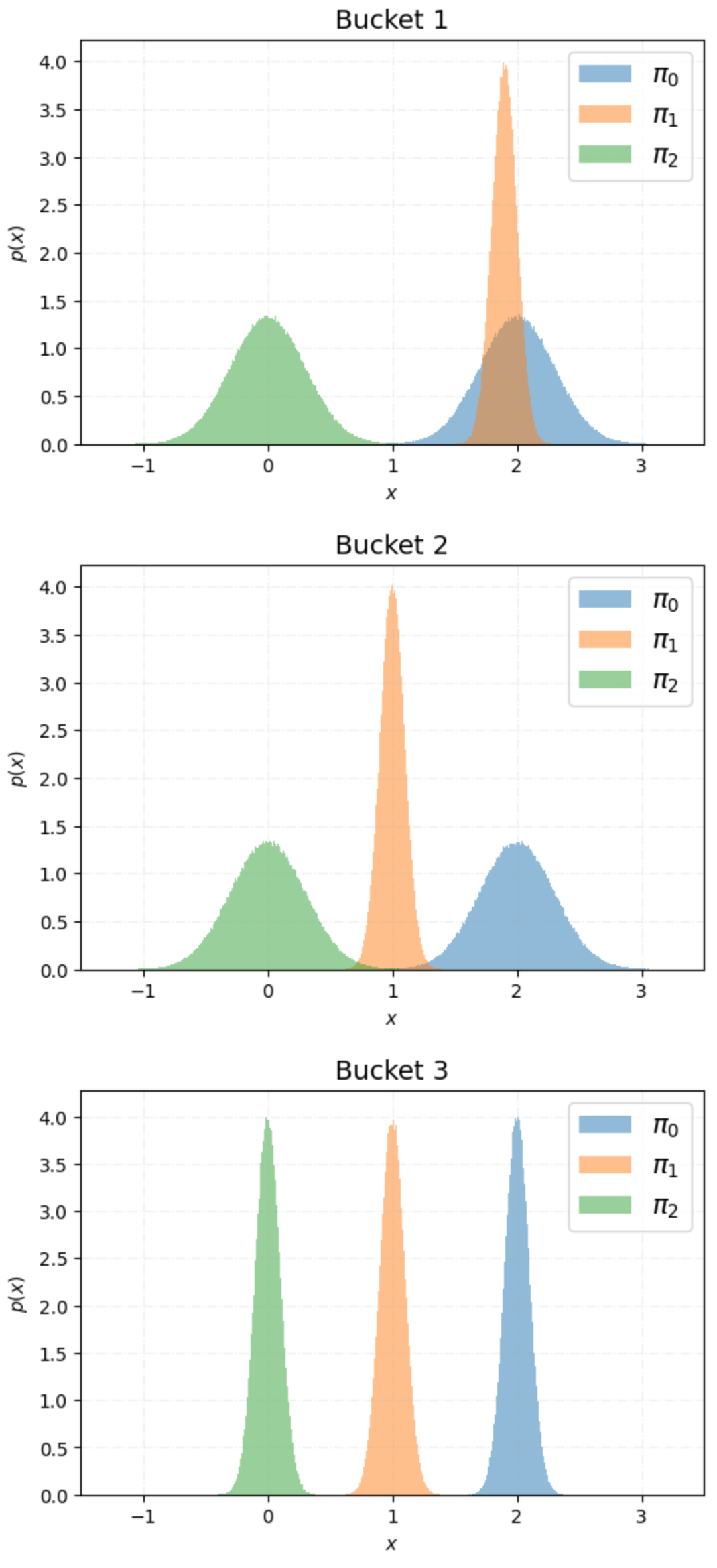}
  \end{center}
  \caption{Synthetic data distributions for small variance case, with parameters described in Table~\ref{table:1d_parameters}.}
  \label{fig:1d_small_var_distr}
\end{figure}

Specifically, we generate parameters of Gaussian distributions for the setting of $M=3$ buckets and $K=3$ policies. We consider two cases: "large variance" and "small variance". The difference is that for "small variance", we scale variances by $0.01$ factor. See Table~\ref{table:1d_parameters} for precise parameters of distributions and Fig.~\ref{fig:1d_large_var_distr} and ~\ref{fig:1d_small_var_distr} for an illustration of policy distributions per bucket. 
For better describing the intuition, let us focus on the first bucket in both cases (first plots of the Figure ~\ref{fig:1d_large_var_distr} and ~\ref{fig:1d_small_var_distr} respectively).

The "large variance" case represents the situation when $\mu_{1,0}=2, \mu_{1,1}=1.9, \Sigma_{1,0}=9, \Sigma_{1,1}=1$, so the difference in means $\mu_{1,0} - \mu_{1,1}$ is much smaller than the difference between variances $\Sigma_{1,0} - \Sigma_{1,1}$. If we assume now $r=0$, it becomes clear that policy $\pi_1$ is one that maximizes the success probability, however the "greedy" approach will choose $\pi_0$ because of the highest mean. 

In the "small variance" case, the relative difference between the variances is now much smaller, meaning that an effective range of $r$, where our algorithm can outperform "greedy", drastically decreases. 

\begin{figure}[h!]
  \begin{center}
    \includegraphics[scale=0.4]{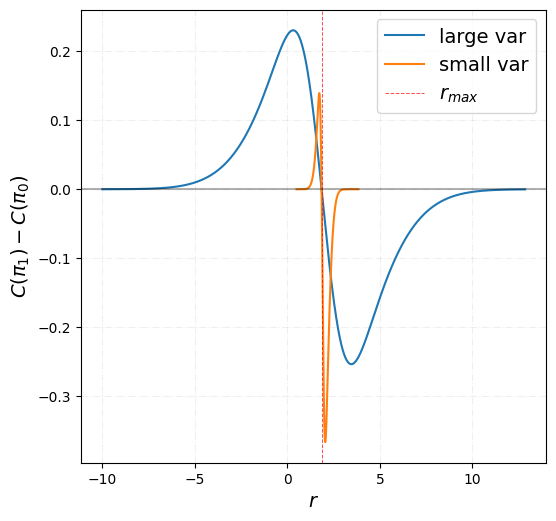}
  \end{center}
  \caption{The difference of criterion values for $\pi_1$ and $\pi_0$.}
  \label{fig:cdfs_diff}
\end{figure}

To illustrate this, we plot in Fig.~\ref{fig:cdfs_diff} the difference of criterion values $$\mathcal{C}(\pi_1) - \mathcal{C}(\pi_0) = \mathbb{P}(Y(\pi_1) > r) - \mathbb{P}(Y(\pi_0) > r)$$ for a range of $r$. 
We define $r_{max}$ as a point where $\rm{cdf}_{\pi_0}(r_{max}) = \rm{cdf}_{\pi_1}(r_{max})$. We can see that $$\mathcal{C}(\pi_1) - \mathcal{C}(\pi_0) \geq 0, r \in [-\infty; r_{max}].$$ The intuition is that while $r$ grows, the left tail of the $\pi_0$ distribution gets outside of the $\mathcal{S}$, then, we reach a point $r_{max}$, where two criterion values are equal. 
Finally, $$\mathcal{C}(\pi_1) - \mathcal{C}(\pi_0) \leq 0, \forall r > r_{max}$$ due to a bigger variance of the $\pi_0$ distribution.

Comparing between the "large" and "small" variance cases, one can clearly see i) a gap in the potential "winning region" size and ii) a difference in the maximum value of $\mathcal{C}(\pi_1) - \mathcal{C}(\pi_0)$.

\subsection{Synthetic setup results}

\begin{figure}
  \begin{center}
    \includegraphics[scale=0.3]{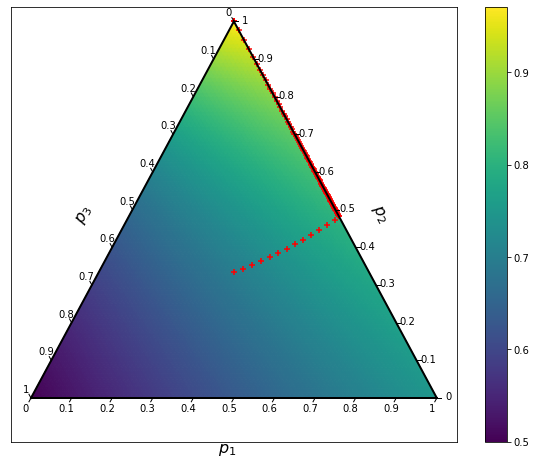}
  \end{center}
  \caption{Convergence of \texttt{SuccessProbaMax} on the toy example. We observe that the optimal decision is different from the output of the greedy algorithm.}
  \label{fig:toy1}
\end{figure}

We use $\mu_{0,k} = [2,1.9,0], \Sigma_{0,k} = [9,1,9], r = 0$ to show the convergence of our algorithm on Fig.~\ref{fig:toy1}. Note that \texttt{SuccessProbaMax} found an optimal allocation $[0,1,0]$, which differs from the one of \texttt{Greedy1D}, $[1,0,0]$. 

To check the algorithm performance where a source of randomness arising from the parameters estimation is presented, we generate normal distributions with parameters $\mu_{g, k}$, $\Sigma_{g, k}$ of sizes $N \in \{1000, 10000\}$ and use \textit{estimations} $\hat{\mu}_{g, k}$, $\hat{\Sigma}_{g, k}$ in the algorithm.

\begin{figure}[h!]
  \begin{center}
    \includegraphics[scale=0.2]{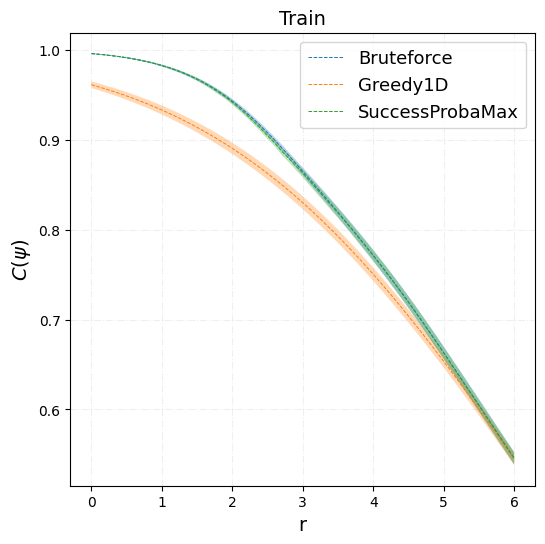}
    \hspace{-0.25cm}
    \includegraphics[scale=0.2]{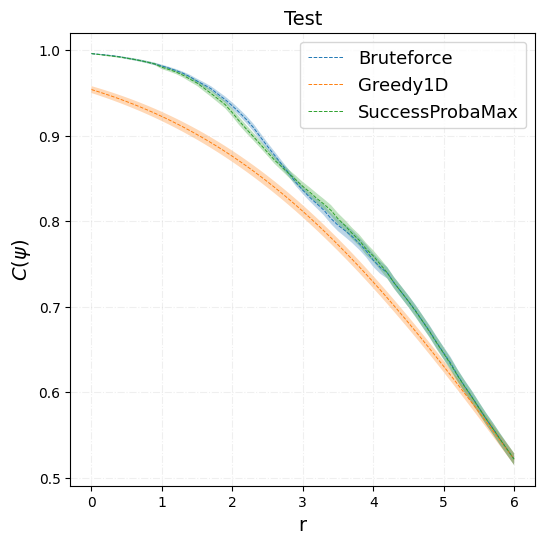}
  \end{center}
  \caption{Synthetic setup with one-dimensional outcome, $N=1000$, "large variance" case: results for a range of difficulty level $r$ on 100 random train (left) and test (right) data splits.}
  \label{fig:large_var_1k_100splits}
\end{figure}

\begin{figure}[h!]
  \begin{center}
    \includegraphics[scale=0.2]{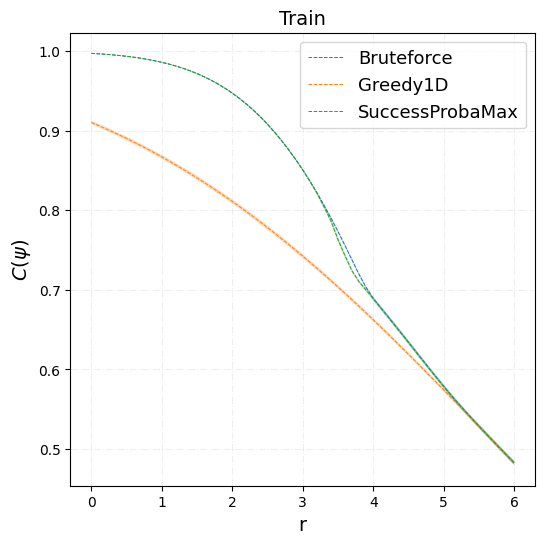}
    \hspace{-0.25cm}
    \includegraphics[scale=0.2]{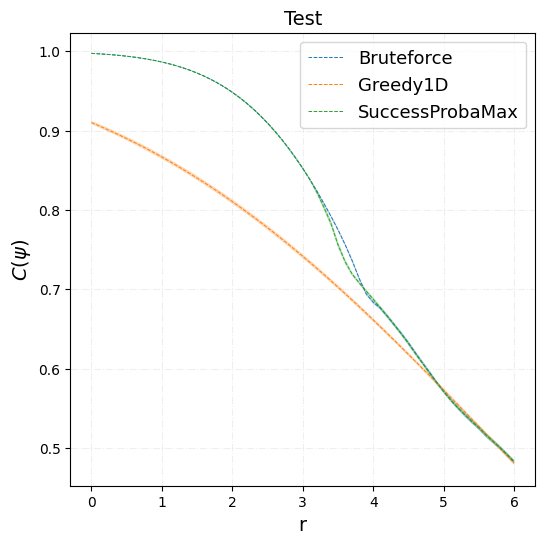}
  \end{center}
  \caption{Synthetic setup with one-dimensional outcome, $N=10000$, "large variance" case: results for a range of difficulty level $r$ on 100 random train (left) and test (right) data splits.}
  \label{fig:large_var_10k_100splits}
\end{figure}

To test the noise coming from the parameters estimation, for both "large" and "small" variance cases we randomly split generated data into train/test parts, estimate $\hat{\mu}_{g, k}$, $\hat{\Sigma}_{g, k}$ on train, and check resulted allocations on both train and test. We repeat the procedure 100 times to build proper confidence intervals. The results for the "large variance" case are presented in Figures ~\ref{fig:large_var_1k_100splits} (for $N=1000$) and ~\ref{fig:large_var_10k_100splits} (for $N=10000$). 

The performance on train and test splits are very similar – it is reasonable as splits contain data from the same distribution. As one can see, the precision of $\hat{\mu}_{g, k}$, $\hat{\Sigma}_{g, k}$ estimation is higher for the $N=10000$ case which is reflected in smaller confidence intervals for each method.

\begin{figure}[h!]
  \begin{center}
    \includegraphics[scale=0.2]{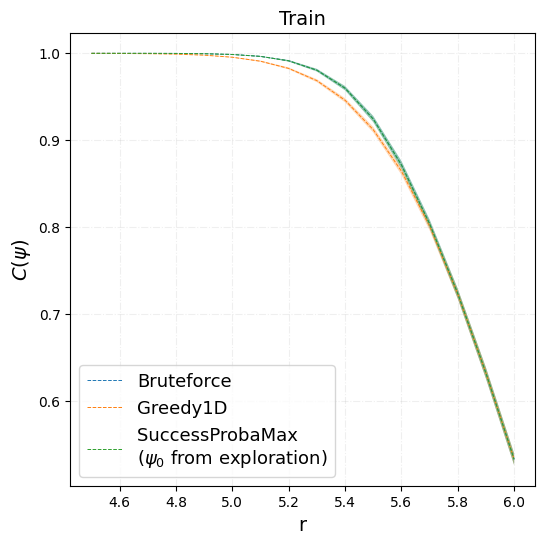}
    \hspace{-0.25cm}
    \includegraphics[scale=0.2]{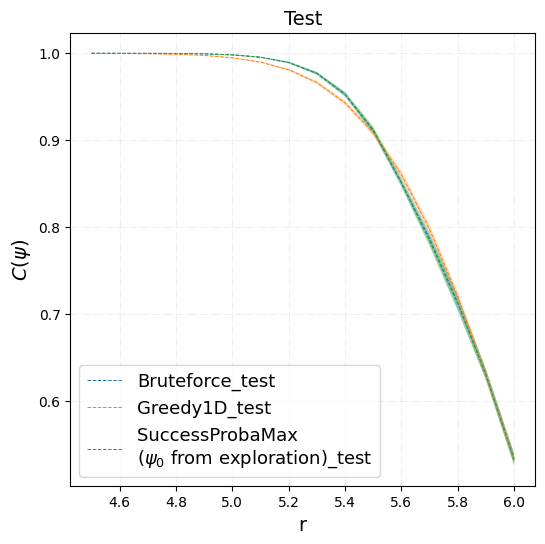}
  \end{center}
  \caption{Synthetic setup with one-dimensional outcome, $N=1000$, "small variance" case: results for a range of difficulty level $r$ on 100 random train (left) and test (right) data splits.}
  \label{fig:small_var_1k_100splits}
\end{figure}

\begin{figure}[h!]
  \begin{center}
    \includegraphics[scale=0.2]{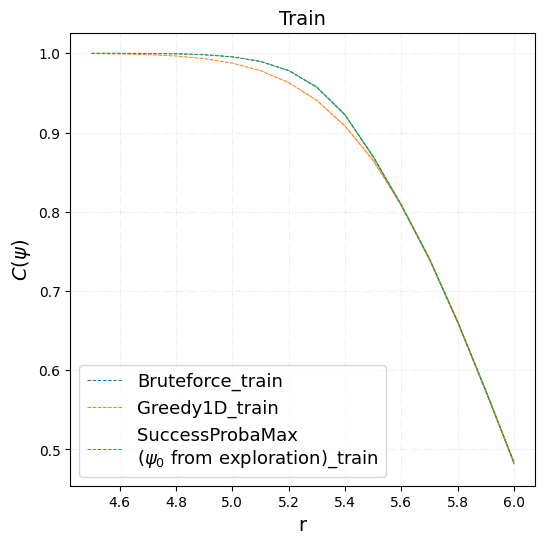}
    \hspace{-0.25cm}
    \includegraphics[scale=0.2]{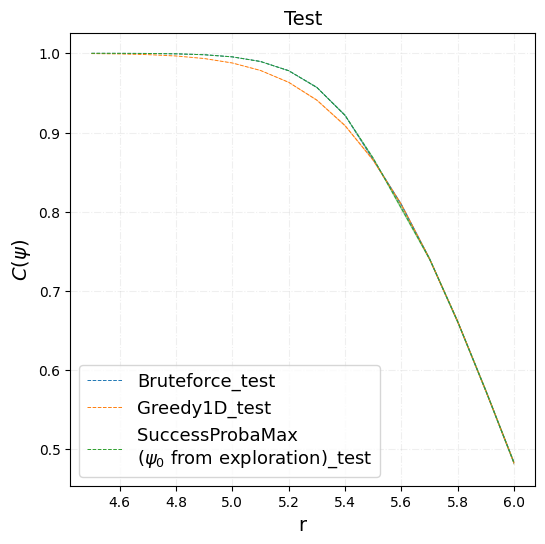}
  \end{center}
  \caption{Synthetic setup with one-dimensional outcome, $N=10000$, "small variance" case: results for a range of difficulty level $r$ on 100 random train (left) and test (right) data splits.}
  \label{fig:small_var_10k_100splits}
\end{figure}

The results for the "small variance" case are provided in Figures ~\ref{fig:small_var_1k_100splits} (for $N=1000$) and ~\ref{fig:small_var_10k_100splits} (for $N=10000$). Due to the smaller original variance, confidence intervals are even smaller than in the previous case.

\subsection{Private dataset}
For the real data cases, it is reasonable to have a direct interpretation of difficulty level $r$ for the RCT success probability. Thus, here we interpret $r$ as a relative gain in the outcome over the reference policy (or simply “Gain" hereafter) that we want to reach. 

\subsection{Private dataset results}

\begin{figure}[h!]
  \begin{center}
    \includegraphics[scale=0.27]{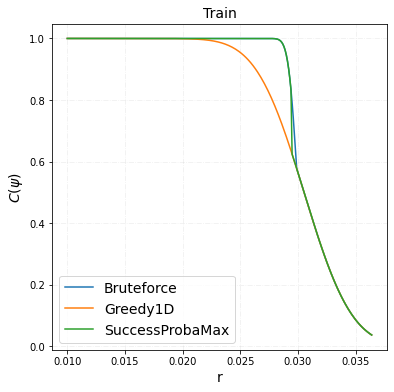}
    \hspace{-0.25cm}
    \includegraphics[scale=0.27]{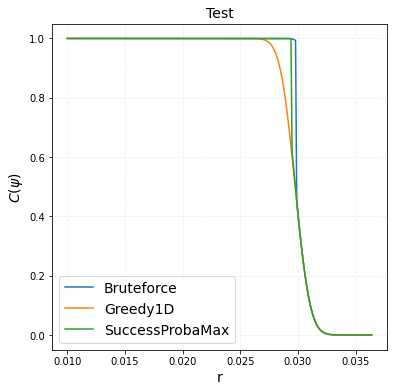}
  \end{center}
  \caption{Results for different Gain $r$ on private data with one-dimensional outcome for train (left) and test (right) splits.}
  \label{fig:priv_1d_res}
\end{figure}

Fig.~\ref{fig:priv_1d_res} describes results on the private dataset with one-dimensional outcome for the range of gains $r$ for train (left) and test (right) splits. Notice that the only possible region to improve $\mathcal{C}(\psi)$ in both cases is $[0.02, 0.03]$. For instance, like \texttt{Bruteforce}, our algorithm initialized with $\psi_0$ from exploration reaches the Gain of 0.029 (2.9\% over the reference) with the probability almost 1, while for \texttt{Greedy1D} it is around 0.7.

\section{Two-Dimensional outcome}
\subsection{Criterion}
In this case we consider an outcome $\mathcal{Y} \in \mathbb{R}^2$, so $\mathbf{Y} = (Y^v, Y^c)$. The problem is parameterized by a two-dimensional difficulty level $\mathbf{r} = (r_v, r_c)$ so that $\mathcal{S}=\{(r_v,+\infty), (-\infty, r_c]\}$. 
The criterion, then, is defined as maximizing the following probability
\begin{align*}
    \EE \left[\mathbb{I}_{\mathcal{S}} \left(\sum_{u\in\mathcal{U}} \mathbf{Y}_u(\psi) \right) \right] = {\rm{cdf}}_{Y^c}(r_c) - {\rm{cdf}}_\mathbf{Y}(\mathbf{r}), 
\end{align*}
where 
\begin{align*}
& \mathbf{Y}(\psi) \sim \mathcal{N}\left(\bmu(\psi), \bSigma(\psi)\right), \\
& \bmu(\psi) = \sum_{k} \sum_{g} \psi(g, k) \bmu_{g,k}, \\
& \bSigma(\psi) = \text{Var}\left[\sum_{k} \sum_{g} \psi(g, k) \mathbf{Y}_{g,k} \right] 
\end{align*}
${\rm{cdf}}_{Y_c}(r_c)$ is the (univariate) c.d.f. of $Y_c$ at $r_c$ and ${\rm{cdf}}_\mathbf{Y}(\mathbf{r})$ is the (bivariate) c.d.f. of $\mathbf{Y}$ at $\mathbf{r}$ \\
\begin{equation}
{\rm{cdf}}_\mathbf{Y}(\mathbf{r}) = \int\limits_{-\infty}^{r_v} \int\limits_{-\infty}^{r_c}f(x_v,x_c) \,dx_v\,dx_c
\end{equation}
where 
$f(x_v,x_c)$ is the  p.d.f. of bivariate normal distribution.

\subsection{Baselines: \texttt{LinProg} and \texttt{MixedInt}}
The \texttt{LinProg} algorithm (linear programming) solves the fractional knapsack problem (soft allocation) and returns a policy with the maximum mean value per bucket under the defined cost constraint. Along with the linear programming approach, we also implement \texttt{MixedInt} (mixed-integer linear programming) that solves the 0/1 knapsack problem and returns a hard allocation.

The main drawback of both algorithms for the success probability maximization problem is that only means of value and cost are used for optimization, while lacking information about the variance.

For both methods, CVXPY Python library~\citep{diamond2016cvxpy} was used for the implementation. 

\subsection{Synthetic setup}
\begin{figure}[h!]
  \begin{center}
    \includegraphics[scale=0.3]{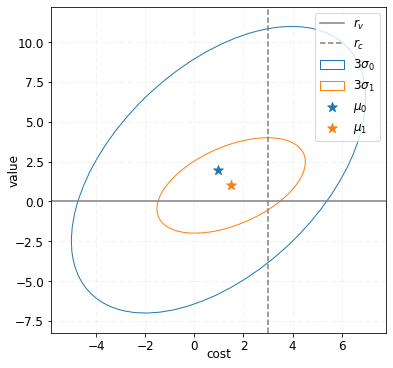}
    \hspace{-0.25cm}
    \includegraphics[scale=0.3]{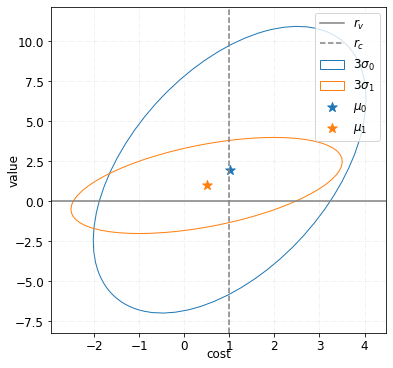}
  \end{center}
  \caption{Confidence ellipses of the synthetic data distributions for case (i) (left) and (ii) (right) with parameters described in Table~\ref{table:2d_parameters}.}
  \label{fig:2d_synth}
\end{figure}

We generate parameters of bivariate Gaussian distributions for the setting of $M = 2$ buckets and $K = 3$ policies. We consider two specific examples, see Table~\ref{table:2d_parameters} for precise parameters of distributions and Fig.~\ref{fig:2d_synth} for an illustration of policy distributions. 

In general, we keep the idea from the one-dimensional setup and create two examples. In example (i), $\pi_1$ has a bigger mean cost and a larger mean value, however, both variances are smaller than for $\pi_0$. For $\mathbf{r} = (r_v, r_c) = (0, 3)$, an optimal policy is $\pi_1$, however, \texttt{LinProg} again will choose $\pi_0$ due to the larger $\mu^v_{1,0}$ at the cost constraint $\mu^c_{1,0}$.

In example (ii), there is a positive difference in means $\mu^v_{1,0} - \mu^v_{1,1}$ but it is much smaller than the difference between variances $\Sigma^v_{1,0} - \Sigma^v_{1,1}$. At the same time, $\mu^c_{1,1}$ is smaller than $\mu^c_{1,0}$ and $\Sigma^c_{1,0} = \Sigma^c_{1,1}$. If we fix $\mathbf{r} = (r_v, r_c) = (0, 1)$, an optimal policy is $\pi_1$, however \texttt{LinProg} will choose $\pi_0$ due to the larger $\mu^v_{1,0}$ at the cost constraint $\mu^c_{1,0}$.


\subsection{Synthetic setup results}
To check the algorithm performance where estimation variability is present, we repeat the same procedure as for the one-dimensional setup, generating bivariate normal distributions of size $N=1000$ with defined parameters. 

\begin{figure}[h!]
  \begin{center}
    \includegraphics[scale=0.3]{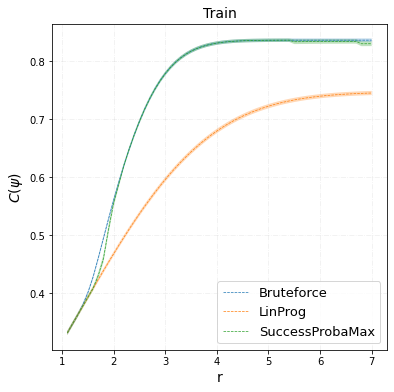}
    \hspace{-0.25cm}
    \includegraphics[scale=0.3]{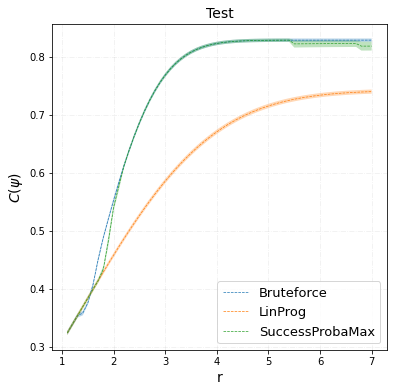}
  \end{center}
  \caption{Synthetic setup with two-dimensional outcome, $N=1000$, case (i): results for a range of $r_c$ on 100 random train (left) and test (right) data splits.}
  \label{fig:2d_i_1k_100splits}
\end{figure}

\begin{figure}[h!]
  \begin{center}
    \includegraphics[scale=0.3]{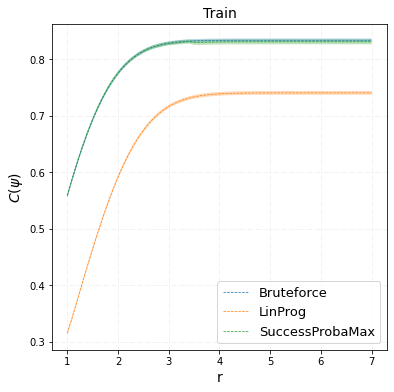}
    \hspace{-0.25cm}
    \includegraphics[scale=0.3]{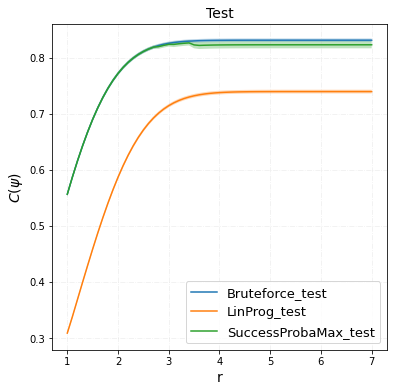}
  \end{center}
  \caption{Synthetic setup with two-dimensional outcome, $N=1000$, case (ii): results for a range of $r_c$ on 100 random train (left) and test (right) data splits.}
  \label{fig:2d_ii_1k_100splits}
\end{figure}

The results for cases (i) and (ii) are presented in Figures~\ref{fig:2d_i_1k_100splits} and~\ref{fig:2d_ii_1k_100splits} respectively. 

\subsection{Private data results}
\begin{figure}[h!]
  \begin{center}
    \includegraphics[scale=0.30]{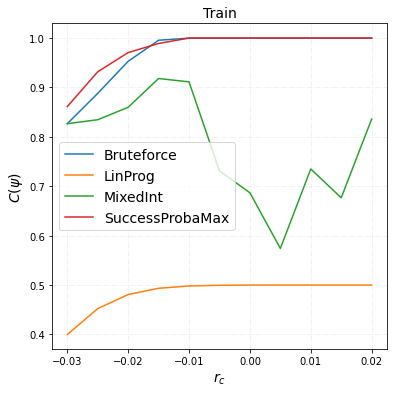}
    \hspace{-0.25cm}
    \includegraphics[scale=0.30]{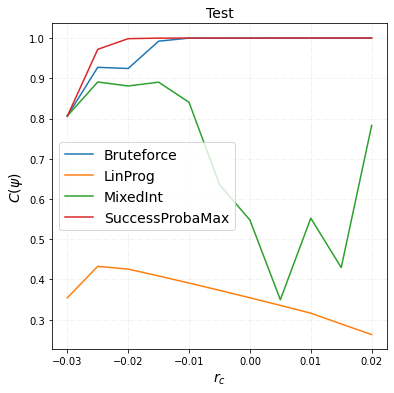}
  \end{center}
  \caption{Results for different Gain $r_c$ while $r_v = 0$ on the private dataset with one-dimensional outcome for train (left) and test (right) splits.}
  \label{fig:priv_2d_res_c}
\end{figure}

Fig.~\ref{fig:priv_2d_res_c} describes results on the private dataset with two-dimensional outcome for the a range of gains $r_c$ while $r_v = 0$ for train (left) and test (right) splits. For instance, \texttt{SuccessProbaMax} reaches the Gain of -0.02 in cost (-2\% over the reference) with probability 0.97 for train and 1 for test, while for \texttt{MixedInt} respective probabilities are 0.86 and 0.89.

\subsection{CRITEO-UPLIFT v2 results}

\begin{figure}[h!]
  \begin{center}
    \includegraphics[scale=0.3]{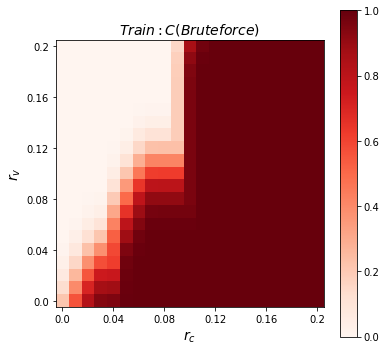}
    \hspace{-0.25cm}
    \includegraphics[scale=0.3]{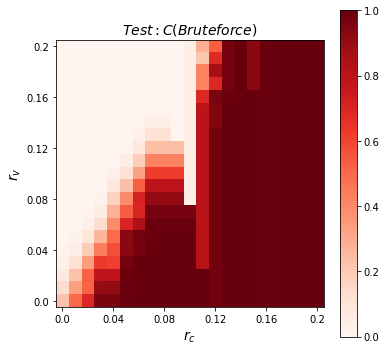}
  \end{center}
  \caption{Absolute criterion values of \texttt{Bruteforce} method on CRITEO-UPLIFT v2 data on train (left) and test (right) data splits.}
  \label{fig:cu2_2d_abs_bf}
\end{figure}

\begin{figure}[h!]
  \begin{center}
    \includegraphics[scale=0.3]{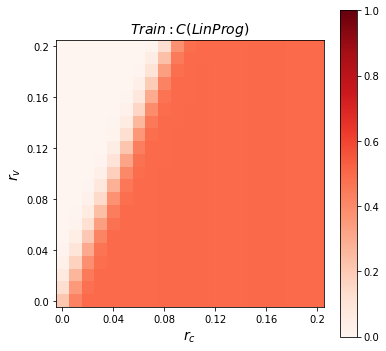}
    \hspace{-0.25cm}
    \includegraphics[scale=0.3]{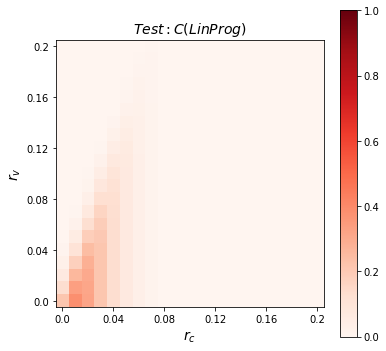}
  \end{center}
  \caption{Absolute criterion values of \texttt{LinProg} method on CRITEO-UPLIFT v2 data on train (left) and test (right) data splits.}
  \label{fig:cu2_2d_abs_lp}
\end{figure}

\begin{figure}[h!]
  \begin{center}
    \includegraphics[scale=0.3]{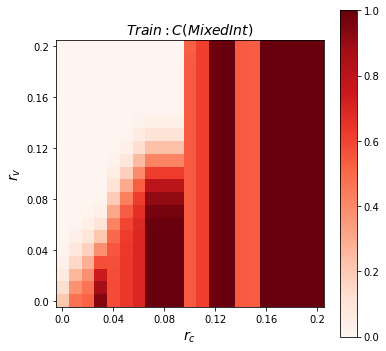}
    \hspace{-0.25cm}
    \includegraphics[scale=0.3]{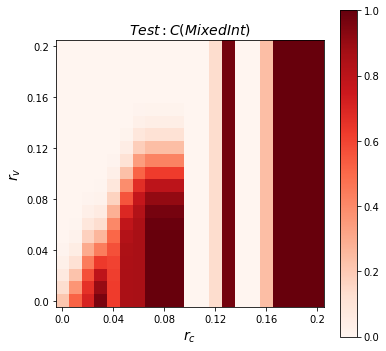}
  \end{center}
  \caption{Absolute criterion values of \texttt{MixedInt} method on CRITEO-UPLIFT v2 data on train (left) and test (right) data splits.}
  \label{fig:cu2_2d_abs_mi}
\end{figure}

\begin{figure}[h!]
  \begin{center}
    \includegraphics[scale=0.3]{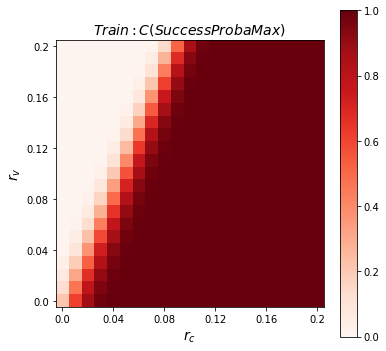}
    \hspace{-0.25cm}
    \includegraphics[scale=0.3]{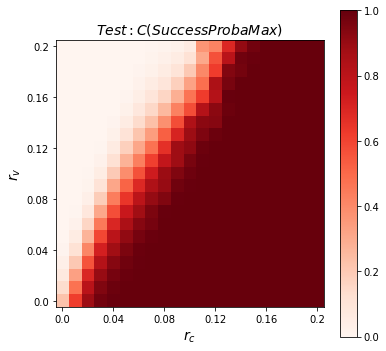}
  \end{center}
  \caption{Absolute criterion values of \texttt{SuccessProbaMax} method on CRITEO-UPLIFT v2 data on train (left) and test (right) data splits.}
  \label{fig:cu2_2d_abs_spm}
\end{figure}

Absolute criterion values for train and test data splits are presented for \texttt{Bruteforce}, \texttt{LinProg}, \texttt{MixedInt} and \texttt{SuccessProbaMax} on Figures~\ref{fig:cu2_2d_abs_bf},~\ref{fig:cu2_2d_abs_lp},~\ref{fig:cu2_2d_abs_mi} and~\ref{fig:cu2_2d_abs_spm} respectively. As we can see, among the other methods our algorithm is the most efficient and stable at the same time.

\section{Hyperparameters}
There are no hyperparameters for the baselines.
\texttt{SuccessProbaMax} includes three hyperparameters - initial allocation $\psi_0$, learning rate $\eta$ and number of steps $n_{st}$. 

For $\psi_0$, we define three options: 
$$\psi_0 \in \{ \psi^{unif}_0, \psi^{baseline}_0, \psi^{expl}_0\},$$
where $\psi^{unif}_0$ represents a uniform allocation, $\psi^{baseline}_0$ is a baseline allocation (from \texttt{Greedy1D} in 1D case and \texttt{LinProg} in 2D), and $\psi^{expl}_0$ is an allocation from exploration, when we generate 50000 random allocations and pick one with the maximum criterion value.

We consider the following possible sets of $\eta$ and $n_{st}$:
$$\eta \in \{10^{-1}, 10^{-2}, 10^{-3}, 10^{-4}\},$$
$$n_{st} \in \{10^4, 10^5, 10^6, 5\cdot 10^6\}$$

For each experiment, we did a grid search over the hyperparameters set aiming to maximize $\mathcal{C}(\psi)$.
The resulted hyperparameters for each experiment are presented in Table~\ref{table:spm_hp}.

\begin{table}[h!]
\caption{Hyperparameters of \texttt{SuccessProbaMax}.}
\label{table:spm_hp}
\centering
\resizebox{0.8\columnwidth}{!}{%
\begin{tabular}{|c|c|c|c|}
\hline
Case   & $\psi_0$ & $\eta$ & $n_{st}$  \\ \hline
1D - synthetic - large var &   $\psi^{unif}_0$    &  $10^{-1}$   &   $10^{4}$   \\ \hline
1D - synthetic - small var &   $\psi^{expl}_0$    &  $10^{-1}$   &   $10^{4}$   \\ \hline
1D - private               &   $\psi^{expl}_0$    &  $10^{-2}$   &   $10^{5}$   \\ \hline
2D - synthetic - case (i)  &   $\psi^{unif}_0$    &  $10^{-2}$   &   $10^{4}$   \\ \hline
2D - synthetic - case (ii) &   $\psi^{unif}_0$    &  $10^{-2}$   &   $10^{4}$   \\ \hline
2D - private               &   $\psi^{expl}_0$    &  $10^{-4}$   &   $5\cdot10^{6}$   \\ \hline
2D - Criteo                &   $\psi^{expl}_0$    &  $10^{-3}$   &   $10^{6}$   \\ \hline
\end{tabular}
}
\end{table}

\section{Hardware}
Experiments were performed on Linux machine with 8 CPUs (Intel(R) Xeon(R) Silver 4108 CPU @ 1.80GHz) and 16Gb of RAM.

\end{document}